\documentclass[preprint]{aastex631}

\usepackage{graphicx}
\usepackage{amsmath,amssymb}
\usepackage{mathrsfs} 
\usepackage{physics}
\usepackage{url}
\usepackage{bm}

\received{xxxx, xxxx}
\revised{xxxx, xxxx}
\accepted{xxxx}

\begin{document}

\title{Probing the mass relation between supermassive black holes and dark matter halos at high redshifts by gravitational wave experiments}

\author{Kazuya Furusawa}
\altaffiliation{furusawa.kazuya.m8@s.mail.nagoya-u.ac.jp}
\affiliation{Department of Physics and Astrophysics, Nagoya University, Chikusa-ku, Nagoya, 464-8602, Japan}

\author{Hiroyuki Tashiro}
\affiliation{Center for Education and Innovation, Sojo University,
Ikeda, Nishi-ku, Kumamoto, 860-0082 Japan}

\author{Shuichiro Yokoyama}
\affiliation{Department of Physics and Astrophysics, Nagoya University, Chikusa-ku, Nagoya, 464-8602, Japan}
\affiliation{Kobayashi-Maskawa Institute for the Origin of Particles and the Universe, Nagoya University, Chikusa-ku, Nagoya 464-8602, Japan}
\affiliation{Kavli IPMU (WPI), UTIAS, The University of Tokyo,\\ 5-1-5 Kashiwanoha, Kashiwa, Chiba 277-8583, Japan}

\author{Kiyotomo Ichiki}
\affiliation{Department of Physics and Astrophysics, Nagoya University, Chikusa-ku, Nagoya, 464-8602, Japan}
\affiliation{Kobayashi-Maskawa Institute for the Origin of Particles and the Universe, Nagoya University, Chikusa-ku, Nagoya 464-8602, Japan}
\affiliation{Institute for Advanced Research, Nagoya University Furocho, Chikusa-ku, Nagoya, 464-8602, Japan}

\begin{abstract}
Numerous observations have shown that almost all galaxies in our Universe host supermassive black holes (SMBHs), but there is still much debate about their formation and evolutionary processes.
Recently, gravitational waves (GWs) have been expected to be a new and important informative observation,
in particular, in the low-frequency region by making use of the Laser Interferometer Space Antenna (LISA) and Pulsar Timing Arrays (PTAs).
As an evolutionary process of the SMBHs, we revisit a dark matter (DM) halo-SMBH coevolution model
based on the halo merger tree employing an ansatz for the mass relation
between the DM halos and the SMBHs at $z=6$.
In this model, the mass of SMBHs grows through their mergers associated with the halo mergers,
and hence, the evolutionary information must be stored in the GWs emitted at the mergers.
We investigate the stochastic gravitational background from the coalescing SMBH binaries, which the PTAs can detect, and also the GW bursts emitted at the mergers,
which can be detected by the mHz band observations such as LISA.
We also discuss the possibility of probing the mass relation between the DM halos and
the SMBHs at high redshift by future GW observations. 
\end{abstract}


\section{Introduction} \label{sec:intro}

Supermassive Black Holes (SMBHs) typically have masses in the range of $10^{6}M_\odot \lesssim M_{\mathrm{BH}}\lesssim 10^{10}M_\odot$ and are found at the centers of almost all observed galaxies. The masses of SMBHs are strongly correlated with various features of the host galaxy, such as velocity dispersion, bulge mass, and luminosity (see, e.g., \citealt{2013ApJ...764..184M}, \citealt{2021MNRAS.507.4274M}). These observations suggest that SMBHs and their host galaxies have co-evolved to date. However, the formation and growth processes of SMBHs are still unknown. To clarify this conundrum, gravitational wave astronomy is expected to provide important clues in the near future through observations of GWs in the low-frequency region with the Laser Interferometer Space Antenna (LISA) \citep{2017arXiv170200786A} and Pulsar Timing Arrays (PTAs) \citep{2013CQGra..30v4009K, 2013CQGra..30v4008M, 2013PASA...30...17M, 2010CQGra..27h4013H}.
Recently, PTA observation collaborations around the world such as NANOGrav\citep{2023ApJ...951L...8A}, EPTA and InPTA \citep{2023arXiv230616214A}, PPTA \citep{2023ApJ...951L...6R}, and CPTA \citep{2023RAA....23g5024X}, have reported 
the evidence of GW background,
which could be basically explained by the superposition of the GWs from SMBH binaries.
These results show that GWs are indeed important for understanding SMBHs, and future precise observations and data analysis are awaited.

There are many theoretical predictions for the GWs from SMBH binaries. The event rate of SMBH coalescence is theoretically estimated by using phenomenological prediction \citep{2003ApJ...583..616J, 2013MNRAS.433L...1S, 2015MNRAS.447.2772R}, semi-analytical models based on the extended Press-Schechter formalism or cosmological N-body simulation \citep{2003ApJ...590..691W,2004ApJ...615...19E,2004ApJ...611..623S, 2009MNRAS.394.2255S,2018MNRAS.477.2599B, 2019MNRAS.483..503Y, 2022MNRAS.509.3488I, 2022A&A...660A..68C}, or cosmological hydrodynamical simulations \citep{2016MNRAS.463..870S,2017MNRAS.471.4508K,2020MNRAS.491.2301K,2020MNRAS.498.2219V}. 

In addition to the relations between SMBHs and various features of host galaxies, there is also a strong correlation between the masses of SMBHs and their host dark matter (DM) halos. For instance, \cite{2002ApJ...578...90F} estimated the local $M_{\mathrm{BH}}$-$M_{\mathrm{halo}}$ relation at $z\sim0$. Furthermore, \cite{2019ApJ...872L..29S} estimated $M_{\mathrm{BH}}$-$M_{\mathrm{halo}}$ relation at $z\sim6$ using 49 Quasi-Stellar Object (QSO) datasets and compared it with the relation at $z\sim0$. 
They argued that these results, together with the relation to galaxy properties, provide clues to understanding how SMBHs grow with galaxies, but the relation at $z=6$ was limited to QSOs with DM halo masses in the range of $10^{11}M_{\odot}\lesssim M_{\mathrm{halo}}\lesssim 10^{13}M_{\odot}$. Their results are thus considered biased toward heavier masses than the most populated mass range $M_{\mathrm{halo}}\lesssim 10^{12}M_{\odot}$.

Using a simple DM halo-SMBH coevolution model based on an extended Press-Schechter formalism that only considers the SMBH mass increase due to mergers, \cite{2023MNRAS.523.3840B} constrained the $M_{\rm BH}$-$M_{\rm halo}$ relation characterized by a power law index $n$ in the region below $10^{12}M_\odot$ and a minimum halo mass $M_{\rm lim}$ that can contain SMBH at $z=6$.
The purpose of this study is to extend their study further and discuss the possibility of restricting the $M_{\rm BH}$-$M_{\rm halo}$ relation by extrapolating from the observed mass range, considering its effect on the GW background radiation from SMBH mergers, and considering the detectability of such GWs in future observations.
In addition to their model, in this study, we also consider the SMBH coalescence timescale and the GWs generated by the coalescence of the SMBH binaries. The $M_{\rm BH}$-$M_{\rm halo}$ relation at $z=6$ allowed from GW observations enables us to constrain the SMBH formation process and its growth by gas accretion at high redshifts $z\gtrsim 6$. 
In particular, the parameter $M_\mathrm{lim}$ may be closely related to the origin of the SMBH. For example, in a scenario where we consider the remnant of Pop.III as the origin of the SMBH \citep{2006NewAR..50...64S,2011ApJ...737...75G}, we can consider a BH of about $10 M_\odot$ in a halo of about $M_{\rm lim}\sim 10^6M_\odot$, and in a direct collapse scenario \citep{2008MNRAS.391.1961D,2016MNRAS.463..529H,2016MNRAS.457.3356V,2019MNRAS.488.3268A} we can consider a BH of about $10^5 M_\odot$ in a halo of about $M_{\rm lim}\sim10^7 M_\odot$. The evolution of these different BH seeds has also been investigated by employing 
semi-analytical models (see e.g. \citep{2022MNRAS.511..616T, 2023MNRAS.518.4672S}).

The structure of this paper is as follows. In Section \ref{sec:DM halo-SMBH}, we explain our DM halo-SMBH coevolution model and calculate the SMBH mass evolution in this model. In Section \ref{sec:GW}, we introduce two types of GW radiation, namely Stochastic GW Background (SGWB) and GW burst, generated by SMBH binaries and discuss their detectabilities and implications to the model parameters $M_{\rm lim}$ and $n$. Finally, in Section \ref{sec:Conclusion}, we conclude our work and discuss future directions.

The calculations in this study adopt the flat $\Lambda\mathrm{CDM}$ model with $h = 0.7$, $\Omega_{\mathrm{m}}=0.3$, $\Omega_{\mathrm{b}}=0.05$, $\Omega_{\mathrm{cdm}}=0.25$, $\Omega_{\Lambda}=0.7$, $n_\mathrm{s}=1$, and $\sigma_8=0.8$. These cosmological parameters are consistent with those in \cite{2019ApJ...872L..29S}.

\section{DM halo-SMBH Coevolution Model} \label{sec:DM halo-SMBH}

The aim of this paper is to investigate the impact of the initial DM halo-SMBH relation
on the SGWB and GW bursts.
As the DM halo-SMBH coevolution model, we take and modify 
the semi-analytical model in~\cite{2023MNRAS.523.3840B},
in which DM halos merge and evolve to satisfy the Press-Schechter mass function 
and only the mergers of SMBHs associated with the DM halo mergers lead to the mass evolution of SMBHs.
Although it is simple, \cite{2023MNRAS.523.3840B} has shown that 
the model can connect the $M_{\rm BH}$-$M_{\rm halo}$ relations
observed between at a high redshift~($z\approx 6$) and the local universe~($z=0$).
In this section, we describe our DM halo-SMBH coevolution model
and the initial DM-SMBH relation at $z=6$.

\subsection{Halo merger tree}\label{subsec:Merger_Tree}

The halo merger tree can provide the DM halo mass evolution in the
hierarchical structure formation described in the Press-Schechter formalism.
There are various methods to construct a halo merger tree~(e.g.,~\citealt{2008MNRAS.389.1521Z}).
In this paper,  we adopt the binary merger tree method first introduced by~\cite{1993MNRAS.262..627L}.
This method is easy to calculate fast and suitable for describing SMBH binary mergers. 
We implement this method in our model following~\cite{1999MNRAS.305....1S}.

In the binary merger tree method, all mergers are binary and
the resultant halo mass after a merger is the sum of 
the two halo masses in the binary merger.
Now we consider the probability of one progenitor halo
with mass $M_{\mathrm{p1}}$ at redshift $z=z_1$ 
evolve to the descendent halo with mass $M_{0}$~($M_{\mathrm{p1}}<M_{0}) $
at $z=z_0$~($z_0<z_1$).
In the extended Press-Schechter formalism~(\citealt{1991ApJ...379..440B}),
such probability is given by
\begin{equation}
\label{EPS_prob}
f(S(M_{\mathrm{p1}}),z_1|S(M_0),z_0)\dd(\Delta S) = \frac{\Delta\delta_c}{\sqrt{2\pi}\Delta S^{\frac{3}{2}}}\exp\Big(-\frac{\Delta\delta_c^2}{2S}\Big)\dd(\Delta S).
\end{equation}
Here $\Delta S \equiv S(M_{\mathrm{p1}})-S(M_0)$ where
$S\equiv\sigma^2(M)$ is the mass variance of the overdensity field.
In Equation~\eqref{EPS_prob},
$\Delta \delta_c \equiv \delta_c(z_1)-\delta_c(z_0)$
with $\delta_c(z) = \delta_c/D(z)$
where $\delta_c = 1.69$ is the threshold of spherical collapse and $D(z)$ is the linear growth factor normalized as $D(z=0)=1$.
This probability corresponds to the one of
the binary merger in which 
the descendant halo with mass $M_0$ forms from
two progenitor halos with masses $M_{\mathrm{p1}}$ and
$M_{\mathrm{p2}}= M_0-M_{\mathrm{p1}}$
in the redshift range $z_0<z<z_1$.
To construct the halo merger tree,
we start from one realization of the halo mass distribution 
following the Press-Schechter mass function at $z=0$.
We take the redshift steps $\Delta z$ up to $z=6$.
At each redshift step, we consider the binary merger probability
in Equation~\eqref{EPS_prob} for each existing DM halo
and obtain two progenitors if the binary merger happens.
We repeat this procedure at the subsequent redshift step until
the redshift reaches $z=6$.
In the construction of the merger tree,
we initially prepare 100 DM halos with mass in the range of $10^{11}M_\odot h^{-1}\leq M_{\mathrm{halo}} \leq 10^{14}M_\odot h^{-1}$ at $z=0$ following the Press-Schechter mass function. 
In the calculation, $\Delta\delta_c$ is often considered constant for reducing 
the calculation cost. In this study, we take $\Delta \delta_c = 0.02$
and we confirmed that the Press-Schechter mass function is realized 
at each redshift step with this value of $\Delta \delta_c$.

Note that
by introducing lower cut-off mass~$M_{\mathrm{lim}}$
we ignore the existence of any DM halo with mass lower than~$M_{\mathrm{lim}}$ in our model.
If a mass lower than $M_{\mathrm{lim}}$ is selected as a progenitor mass in the merger tree construction,
we assume that this mass is accreted onto the other halo progenitor as diffuse DM, and we do not consider it as a DM halo.
We will also discuss $M_{\mathrm{lim}}$ in Section~\ref{subsec:Our_Model}.

\subsection{Initial mass relation between SMBHs and DM halos}

\cite{2019ApJ...872L..29S} has reported the mass relation between
the DM halos and SMBHs around $z=6$ using QSO datasets.
Motivated by this work, 
\cite{2023MNRAS.523.3840B} suggested the power-law type of the 
mass relation between
the DM halos and SMBHs at $z=6$,
\begin{equation}
\label{init_powerlaw}
\frac{M_{\mathrm{BH}}}{M_1} = \Big(\frac{M_{\mathrm{halo}}}{M_0}\Big)^n~~{\rm for}~M_{\rm halo} \geq M_{\rm lim},
\end{equation}
where $(M_0,M_1)=(3.5\times 10^{12}M_\odot,1.5\times 10^9M_\odot)$. 
We adopt this power-law type of the mass relation characterized by a power-law index~$n$ 
as the initial mass distribution of SMBHs in our calculation.
As introduced in the previous section, 
$M_{\rm lim}$ is the lowest mass of DM halos 
and, in our merger tree, the increment of mass lower than $M_{\rm lim}$
is regarded as the accretion of ambient DM, which does not host an SMBH. 

\subsection{DM halo-SMBH coevolution}\label{subsec:Our_Model}

Based on the halo merger history from $z=6$ to $z=0$ and the initial condition above, we consider the evolution of SMBH via SMBH mergers accompanied by their host DM halo mergers. 
Generally, the SMBH coalescence is delayed from the mergers of DM halos. In order to take into account the delay, we adopt the SMBH merger scenario following~\cite{2003ApJ...582..559V}.

At a merging time $t_\mathrm{mg}$, a lighter DM halo is captured into a heavier one. Then the lighter DM halo sinks to the center of the heavier DM halo due to dynamical friction.
Although the shape of the lighter DM halo decays during the dynamical friction process, 
the lighter DM halo can keep its central SMBH.
Therefore, when the lighter DM halo reaches the center of the heavier DM halo,
we can assume that the coalescence between SMBHs that each DM halo hosts occurs. 
Hence we can take this dynamical friction timescale $t_\mathrm{df}$ as the SMBH merger timescale\footnote{Recently, \cite{2023MNRAS.523..758C} investigated the merger rate by using cosmological hydrodynamical simulation adopting the dynamical friction timescale between a lighter BH and the stellar component in the host galaxy having the heavier BH at the center. In Appendix~\ref{sec:App.A}, we will discuss the model by using this dynamical friction timescale. 
There is, in fact, a lot of discussion about the timescale of SMBH mergers, which is related to the so-called 'final parsec problem' (see e.g. \cite{2014SSRv..183..189C} for the detailed SMBH merger scenario). Also, we will additionally consider the effect of the hardening phase in Appendix~\ref{sec:App.B}.}. We calculate $t_\mathrm{df}$ as

\begin{equation}
\begin{split}
t_{\mathrm{df}} &= \frac{1.17}{\ln\Lambda}\frac{r_\mathrm{i}^2 V_\mathrm{circ}}{GM_{\mathrm{L}}}\epsilon^\alpha = \frac{1.17}{\ln\Lambda}\frac{r_{\mathrm{vir}}^2 V_\mathrm{circ}}{GM_{\mathrm{L}}}\Theta\\
&=0.731\mathrm{Gyr}\frac{1+P}{P\ln{(1+P)}}\Big[\frac{\Omega_{\mathrm{m}}}{\Omega_\mathrm{m}^z}\frac{\Delta_{\mathrm{vir}}}{18\pi^2}\Big]^{-\frac{1}{2}}(1+z)^{-\frac{3}{2}},\\
\end{split}
\label{tdf}
\end{equation} 
where $P=M_\mathrm{L}/M_\mathrm{H}$ is the mass ratio of the lighter-heavier DM halos ($M_\mathrm{L}\leq M_\mathrm{H}$), $V_\mathrm{circ}$ is the circular velocity of the new halo with mass $M_\mathrm{L} + M_\mathrm{H}$, $r_{\mathrm{vir}}$ is the virial radius, and $\ln\Lambda \simeq \ln(1+P)$ is the Coulomb logarithm. We calculate $V_{\mathrm{circ}}$ and $r_{\mathrm{vir}}$ following \cite{2001PhR...349..125B}, where $\Delta_{\mathrm{vir}} = 18\pi^2 + 82(\Omega_\mathrm{m}^z-1) - 39(\Omega_\mathrm{m}^z-1)^2$ and $\Omega_{\mathrm{m}}^z = \Omega_\mathrm{m}(1+z)^3/[\Omega_{\mathrm{m}}(1+z)^3+\Omega_\Lambda]$. This timescale depends on $\Theta$ which is introduced to represent 
the initial orbit of the satellite and defined as $\Theta = (r_\mathrm{i}/r_{\mathrm{vir}})^2\epsilon^\alpha$. Here $r_\mathrm{i}$ is the radius of the circular orbit having the same energy as the actual non-circular orbit of the satellite halo, $\epsilon$ is the circularity of the actual orbit, and $\alpha$ is the parameter to fit the timescale to the results of numerical simulations (see, e.g., \citealt{1999ApJ...525..720C}, \citealt{2003ApJ...582..559V} and \citealt{2003MNRAS.341..434T}). Here, we choose the set of constants in Equation 3,  $\Theta = 0.3$, following Volonteri et al. (2003)
Then the coalescence of the SMBH binary occurs at time $t_\mathrm{mg} + t_\mathrm{df}$ after the DM halo merger at time $t_\mathrm{mg}$. The mass of the final SMBH equals the total mass of two SMBHs. 
\begin{figure}[t!]
\centering
\includegraphics[width=0.6\textwidth]{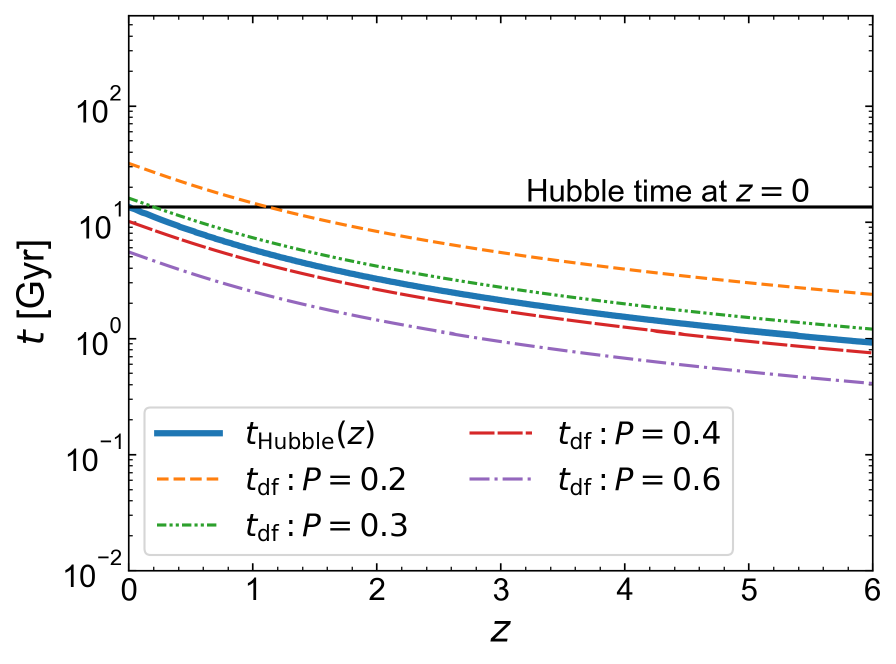}
\caption{Dynamical friction timescale $t_\mathrm{df}$ 
as a function of redshift $z$, which is given by Equation~\eqref{tdf}. 
The thick solid line (blue) shows the Hubble time 
as a function of $z$ and the thin solid line (black) corresponds to the Hubble time at $z=0$.
We plot the timescale for different mass ratios of two merging halos, $P=0.2$, $0.3$, $0.4$, and $0.6$ from top to bottom. }
\label{fig:tdf_HH}
\end{figure}
\begin{figure}[t!]
\centering
\includegraphics[width=0.6\textwidth]{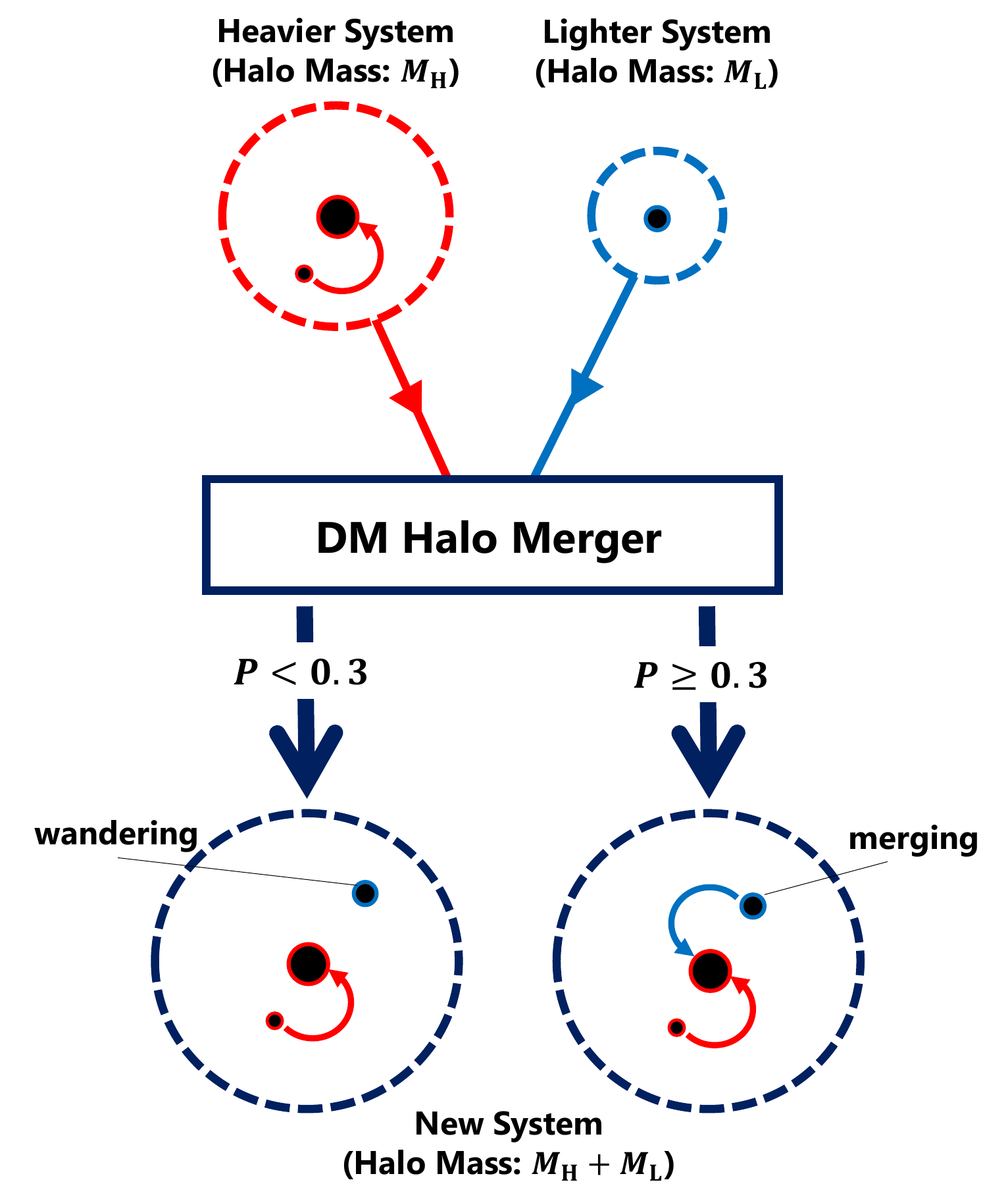}
\caption{Illustration for the merger of two systems~(DM halos with SMBHs) in our model. The heavier system consists of a DM halo with mass $M_{\rm H}$~(a dashed circle in red) and
two SMBHs (black-filled circles with red outlines).
One SMBH is at the center of the DM halo, and the other SMBH is falling to the center.
The lighter system includes a DM halo with mass $M_{\rm L}$~(a dashed circle in blue)
and an SMBH at the center (black-filled circles with blue outlines).
Through the DM halo merger, a new system has a DM halo with mass $M_{\mathrm{H}}+M_{\mathrm{L}}$
(a dashed circle in navy) and two SMBHs coming from the heavier system~(one is at the center and the other is falling to the center).
If $P<0.3$, we neglect the SMBH from the lighter system in the new system (the left branch).
Otherwise, we assume that the lighter system is falling into the center of the new system
with the dynamical friction timescale determined by $P$ and
leads to the SMBH coalescence.}
\label{fig:DMhalo-SMBH Merger}
\end{figure}

Figure~\ref{fig:tdf_HH} shows the dynamical friction timescale for each z with different values of $P$. As shown in this figure, the SMBH merger timescale is determined by $P$. Also, this figure finds out that the dynamical friction timescale has almost the same dependence on $z$ as the Hubble time. Therefore, here we introduce the critical value~$P_*$ with which the dynamical friction timescale is roughly the same as the Hubble timescale at time~$t_\mathrm{mg}$. If~$P$ of the merging DM halos is smaller than~$P_*$, the dynamical timescale exceeds the cosmological timescale and the coalescence of SMBHs does
not happen in the cosmological timescale. 
Therefore, we neglect the SMBH coalescence with $P<P_*$. 
If $P$ is larger than $P_*$, the dynamical timescale is short enough to 
induce the SMBH coalescence in the cosmological time.
Therefore, we assume that the SMBH coalescence with $P\geq P_*$
leads to the mass growth of the central SMBH of the heavier DM halo
at time $t_{\rm mg} + t_\mathrm{df}$.
In our redshift range, we confirmed that the dynamical friction timescale is always
less than the Hubble timescale with $P \geq 0.3$.
The DM halo merger with $P \geq 0.3$ corresponds to the so-called major merger. 
Through our simulations, we adopt the constant critical value 
$P_* = 0.3$.

In our model, the SMBH coalescence is delayed from the DM merger.
Therefore, there is the possibility that, 
before the satellite SMBH coalesces with the central SMBH, 
the host DM halo has a merger with another DM halo. 
As shown in Figure~\ref{fig:DMhalo-SMBH Merger},
let us suppose that 
the heavier DM halo (red dashed circle) has one central SMBH and one satellite SMBH, 
while the lighter DM halo (blue dashed circle) has one central SMBH.
After the DM halo merger at time $t_\mathrm{mg}$, the merged DM halo (navy) succeeds the central and satellite SMBHs in the heavier DM halo of the progenitors (red). 
If $P=M_\mathrm{L}/M_\mathrm{H} \geq 0.3$, 
we regard the central SMBH in the lighter DM halo as the satellite SMBH in the merged DM halo which 
eventually coalesces with the central SMBH
at $t_\mathrm{mg}+t_\mathrm{df}$ (the right branch in Figure~\ref{fig:DMhalo-SMBH Merger})
\footnote{If a satellite SMBH is also present in the lighter DM halo and the conditions for merging with the central SMBH of that halo are satisfied by time $t=t_{\rm mg}+t_{\rm df}$, then we consider that the two SMBHs have merged.}.
While, if $P=M_\mathrm{L}/M_\mathrm{H} <0.3$, we neglect the merger between the central SMBHs of the two DM halos because they do not coalesce in the cosmological timescale (the left branch in Figure~\ref{fig:DMhalo-SMBH Merger}). 

\begin{figure}[t!]
\centering
\includegraphics[width=0.6\textwidth]{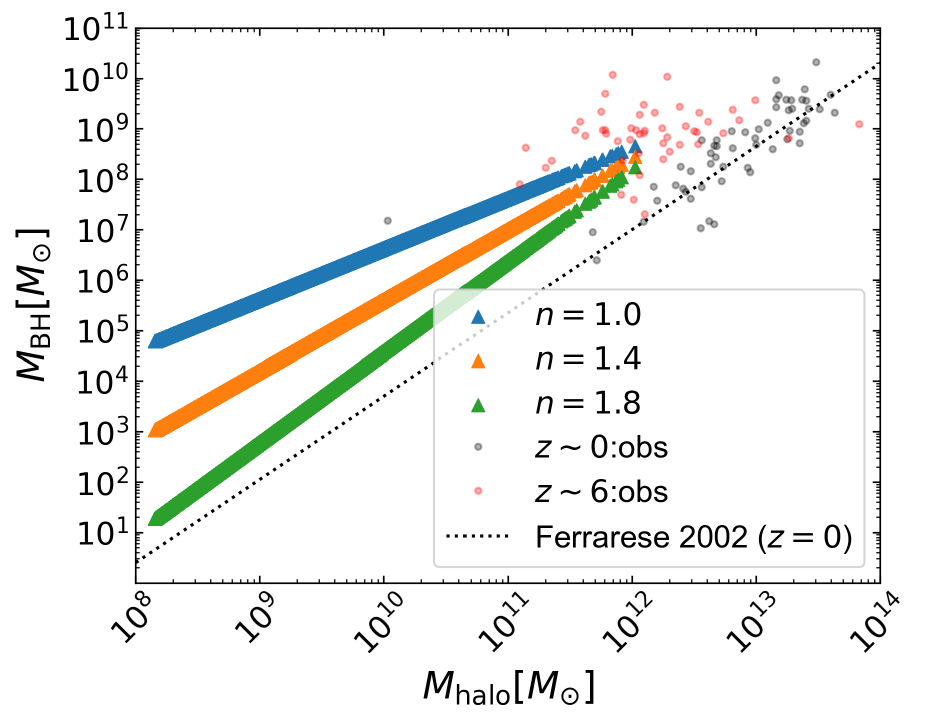}
\caption{The mass relation between DM halo and SMBHs at $z=6$.
We plot our initial mass relation obtained from the halo merger tree and Equation~\eqref{init_powerlaw} with $M_{\mathrm{lim}}=1.0\times 10^8M_{\odot}h^{-1}$
as colored triangles.
The different color represents the different power-law index~$n$ in 
Equation~\eqref{init_powerlaw}; $n=1.0$ in blue,  $n=1.4$ in orange
and  $n=1.8$ in green.
The faint black and red small circles represent
the observed mass relation at $z=0$ and $z=6$~\citep{2019ApJ...872L..29S}, respectively.
We also plot the local relation given by~\cite{2002ApJ...578...90F}
in a black dotted line.}
\label{fig:HaloBHz6_108}
\end{figure}

Now we show the results of the SMBH mass growth in our model.
Figure \ref{fig:HaloBHz6_108} represents the initial mass relation between the masses of DM halos and SMBHs at $z=6$
with different power-law indexes in our calculations. The mass distribution of DM halos is obtained 
from the procedure we have described above.
Then we assigned SMBHs to DM halos at $z=6$, following Equation~\eqref{init_powerlaw}.
For comparison, we also plot the observed mass relation between DM halos and SMBHs at $z=6$ 
and $z=0$ as in red and gray dots, respectively. 
We also represent the local mass relation reported in~\cite{2002ApJ...578...90F} as 
the black dotted line.
Some of the DM halo masses among observed DM halo-SMBH systems are larger than $10^{12}~M_\odot$ at $z=6$, while the main mass range of DM halos in our simulations is less than $10^{12}~M_\odot$.

\begin{figure}[t!]
  \centering
    \centering
    \includegraphics[width=0.9\textwidth]{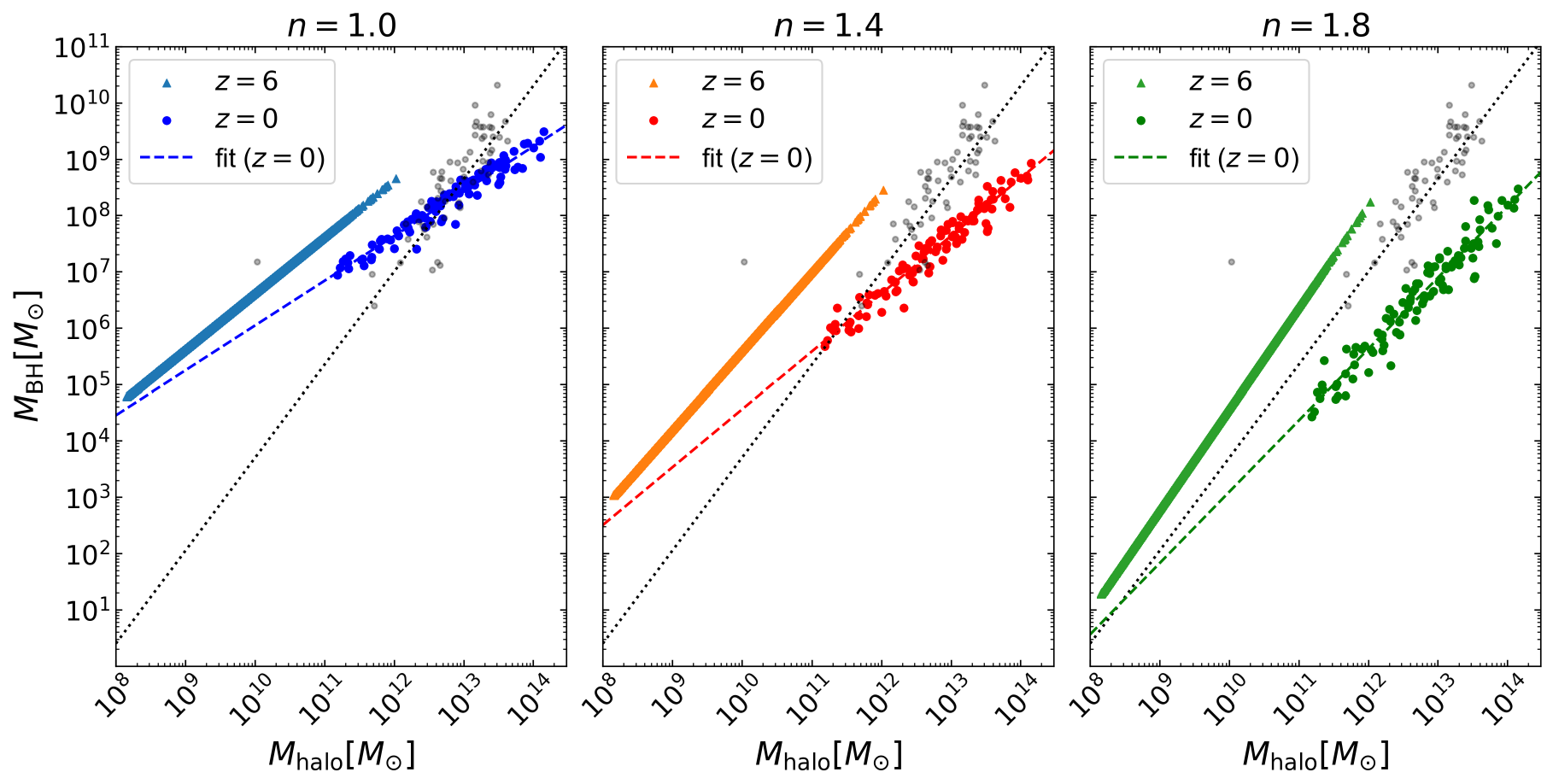}
    \centering
    \includegraphics[width=0.9\textwidth]{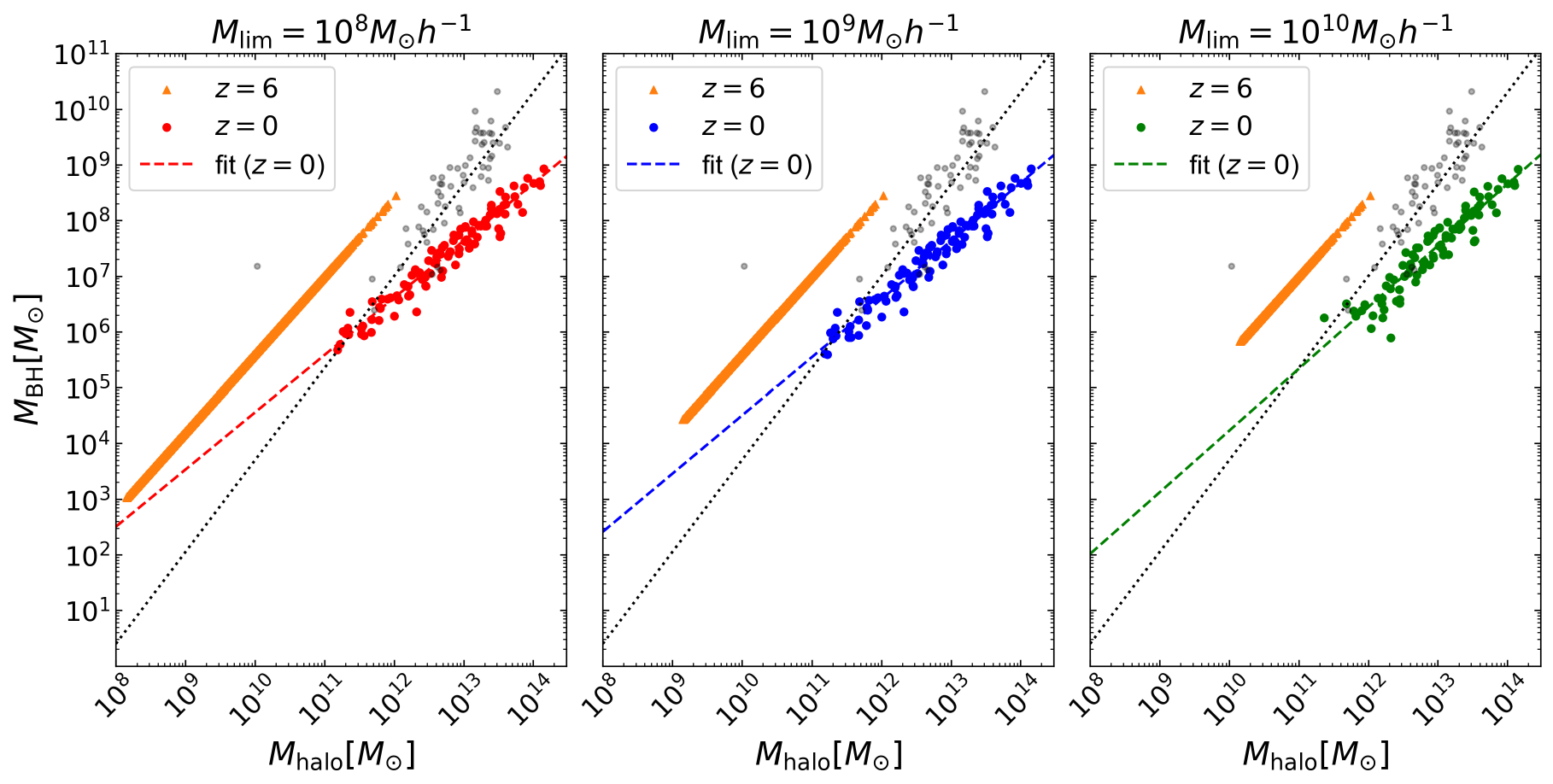}
\caption{The mass relation between DM halos and SMBHs at $z=0$. 
The upper three panels show the dependence on the power-law index $n$ with the
fiducial $M_{\mathrm{lim}}=1.0\times 10^8M_{\odot}h^{-1}$ and the lower three panels show the dependence on the minimum halo mass $M_{\mathrm{lim}}$ with $n=1.4$.
In each panel, colored circles represent the mass relations in the simulation 
and the colored dashed line is the fitting power-law line for 
the simulation result at $z=0$. 
The faint small black circle and the black dotted line are 
the same as in Figure~\ref{fig:HaloBHz6_108}. 
Also, in the all panels, we plot the initial mass relation at $z=6$ with the colored triangles.
}
\label{fig:HaloBHz0}
\end{figure}

Figure~\ref{fig:HaloBHz0} tells us the resultant mass relation between
DM halos and SMBHs at $z=0$ from the initial relation in Figure~\ref{fig:HaloBHz6_108}.
In the left panel of Figure~\ref{fig:HaloBHz0}, we show the dependence on the power-law
index~$n$ with the fiducial $M_{\rm lim}$.
Although all of DM halo mass is below $10^{12} M_\odot$ at $z=6$ initially, they evolve through the mergers and accretions in each timestep. Resultantly, the mass range of DM halos expands to $10^{14} M_\odot$ at $z=0$. On the other hand, the SMBH mass grows only through the major mergers of DM halos with $P>0.3$ in our model.  Therefore, the mass growth rate of SMBH is lower than that of DM halo. 
Because of the slower mass growth of SMBH, the power-law index $n$ in the mass relation between DM halos and SMBHs becomes small as the redshift goes down. We find out that $n=1.0$, $1.4$, and $1.8$ at $z=6$ shift to $0.9$, $1.1$, and $1.3$ at $z=0$, respectively. The more effective growth is required to maintain the larger power-law index. As a result, the difference of the power-law indices from $z=6$ and $z=0$ becomes larger for the steeper initial power-law index at z=6. 

In the lower panel of Figure~\ref{fig:HaloBHz0}, 
we plot the mass relations for different $M_{\rm lim}$.
In contrast to $n$, the difference of $M_{\rm lim}$
does not affect the evolution of the mass relation.
Although changing $M_{\rm lim}$ just modifies the 
total SMBH mass, it does not change 
the mass dependence of the SMBH mass growth. 

At the last of this section, it is worth mentioning 
about the SMBH growth via gas accretions.
Although we omit it in our model for simplicity, 
the gas accretion onto an SMBH is another major contribution 
for the mass growth of SMBHs.
Therefore, we underestimate the final SMBH mass at $z=0$.
Including the mass growth via the gas accretion shifts up
the SMBH mass $z=0$ in the all of mass range.
Although there exists the gap between the observed local relations
and the relations in our simulations in Figure~\ref{fig:HaloBHz0},
the gas accretion onto SMBHs can decrease this gap.

\section{Gravitational Waves from SMBH Binaries in the Coevolution Model} \label{sec:GW}

Based on the DM halo-SMBH coevolution model introduced in the previous section, we investigate GWs from SMBH binaries in the process of coalescing evolutions. First, we calculate the SGWB that could be mainly produced by the superposition of a large number of inspiraling phases and discuss the constraint from the PTA experiments on the power-law index, $n$, which has been introduced in Equation~\eqref{init_powerlaw}.
Then, we also estimate the expected event rate of SMBH mergers in the LISA experiment.

\subsection{Stochasic gravitational wave background}

To describe the energy density of the SGWB per logarithmic frequency interval, $\dd\rho_{\rm GW}/\dd\ln f$, we use the characteristic strain amplitude $h_c(f)$ which is given as (see, e.g., \citealt{Maggiore:2018sht})
\begin{equation}
\label{def_OmegaGW}
\frac{\dd\rho_{\mathrm{GW}}}{\dd\ln f} = \frac{\pi c^2}{4G}f^2h_c^2(f),
\end{equation}
and the SGWB from SMBH binaries can be estimated as
\begin{equation}
\label{OmegaGW_SMBH}
\begin{split}
&\frac{\dd\rho_{\mathrm{GW}}}{\dd\ln f}=\int{\dd z\dd M_1 \dd M_2} \frac{\dd^3n_c(z,M_1,M_2)}{\dd z \dd M_1 \dd M_2} \, \frac{\dd E_{\mathrm{GW}}(f,z,M_1,M_2)}{\dd \ln f}~.\\
\end{split}
\end{equation}
Here, $\dd^3 n_c / \dd z \dd M_1 \dd M_2$ is the comoving number density of the coalescing SMBH binaries with masses $M_1\sim M_1+\dd M_1$ and $M_2\sim M_2+\dd M_2$ at $z \sim z + \dd z$, which can be calculated in the DM halo-SMBH coevolution model discussed in the previous section,
and $\dd E_{\mathrm{GW}}(f,z,M_1,M_2) / \dd \ln f$ is the energy spectrum of the GW radiation from a coalescing SMBH binary with masses of $M_1$ and $M_2$ at the redshift $z$.
As the characteristic GW frequency from the BH binary, we introduce $f_{ISCO}$, which is the frequency of GWs from the BH binary with the binary separation corresponding to the innermost stable circular orbit (ISCO). In the observed frame, $f_{ISCO}$ for the BH binary at a redshift $z$ is given by
\begin{equation}
\label{f_ISCO}
f_{\mathrm{ISCO}} = \frac{c^3}{6^{\frac{3}{2}}\pi G M_{\mathrm{tot}}}\frac{1}{1+z} = 4.7\times10^{-7}~\mathrm{Hz}\Big(\frac{1}{1+z}\Big)\Big(\frac{10^{10}M_{\odot}}{M_{\mathrm{tot}}}\Big),
\end{equation}
where $M_{\mathrm{tot}} = M_1 + M_2$.
For the lower frequencies than the $f_{\rm ISCO}$ the coalescing BH binaries can be considered to be in the inspiral phase, and
the most sensitive frequency range of the PTA experiments is about $10^{-9} \sim 10^{-8}~\mathrm{Hz}$.
As we have shown in the previous section, the mass range of the SMBHs in our setup is typically $10^{4} \sim 10^{10} M_\odot$ (see Figures.~\ref{fig:HaloBHz6_108} and \ref{fig:HaloBHz0}).
Thus, for the PTA experiments, we can use the energy spectrum of the GW radiation in the inspiraling phase,
which is given by (see, e.g., \citealt{Maggiore:2018sht})
\begin{equation}
\label{spec_inspiral}
\frac{\dd E_{\mathrm{GW}}}{\dd \ln f}(f,z,M_1,M_2) = \frac{1}{3G}(GM_c)^{\frac{5}{3}}\Big(\frac{\pi f}{1+z}\Big)^{\frac{2}{3}}~~\left({\rm for}~f < f_{\mathrm{ISCO}} \right)~.
\end{equation}
where $M_c=(M_1 M_2)^{\frac{3}{5}}/(M_1+M_2)^{\frac{1}{5}}$ is the chirp mass of the binary. For simplicity, we have assumed the circular orbit and $\dd E_{\mathrm{GW}}/\dd \ln f = 0$ for $f>f_{\rm ISCO}$.
Then, from the above formula, we can calculate
the characteristic strain amplitude of the SGWB from SMBH binaries
in our DM halo-SMBH coevolution model 
and it basically depends on two model parameters, $n$ and $M_{\rm lim}$, which have been introduced in the ansatz
given by Equation~\eqref{init_powerlaw}.

\begin{figure}[t!]
  \centering
    \centering
    \includegraphics[width=0.9\textwidth]{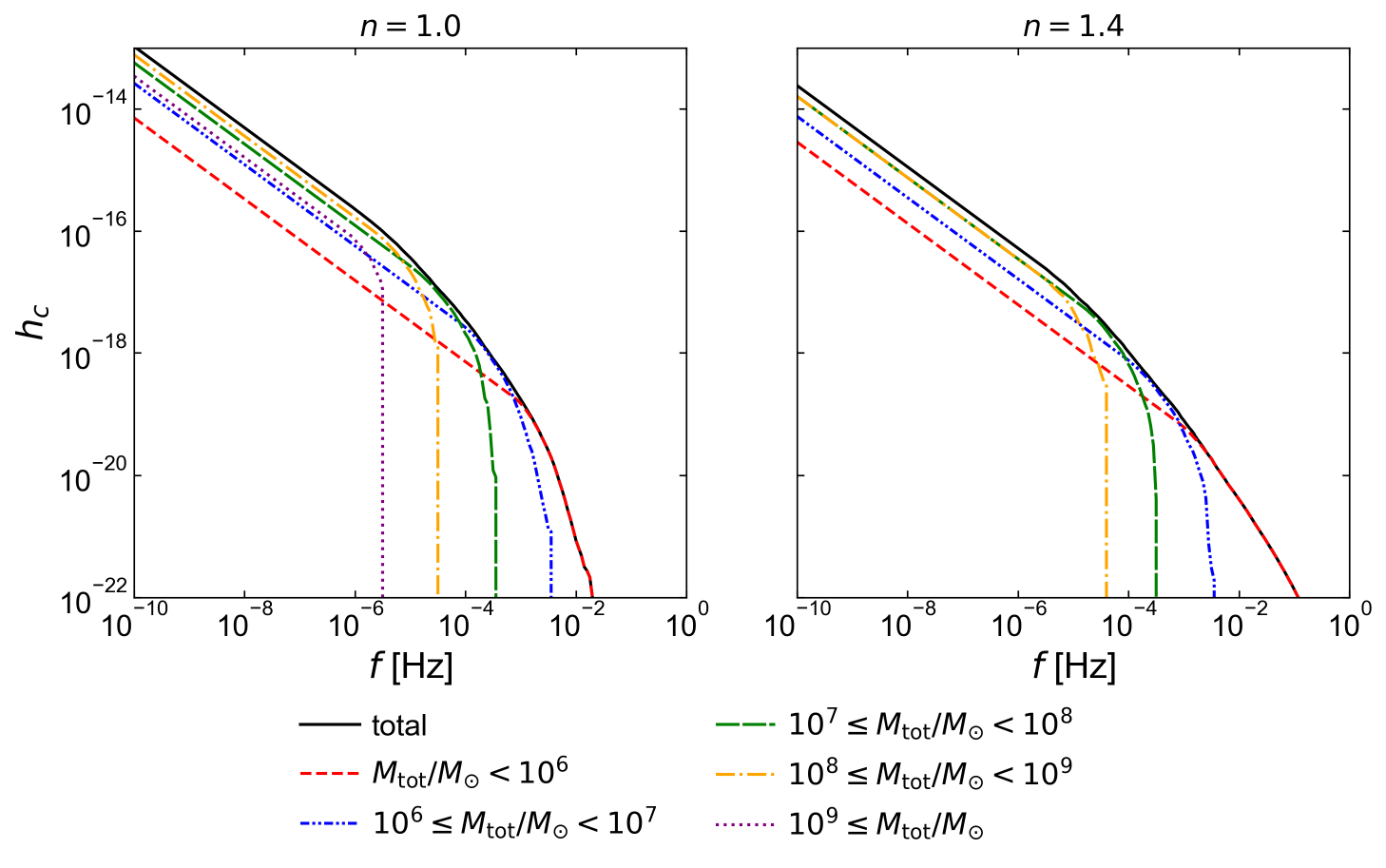}
    \centering
    \includegraphics[width=0.9\textwidth]{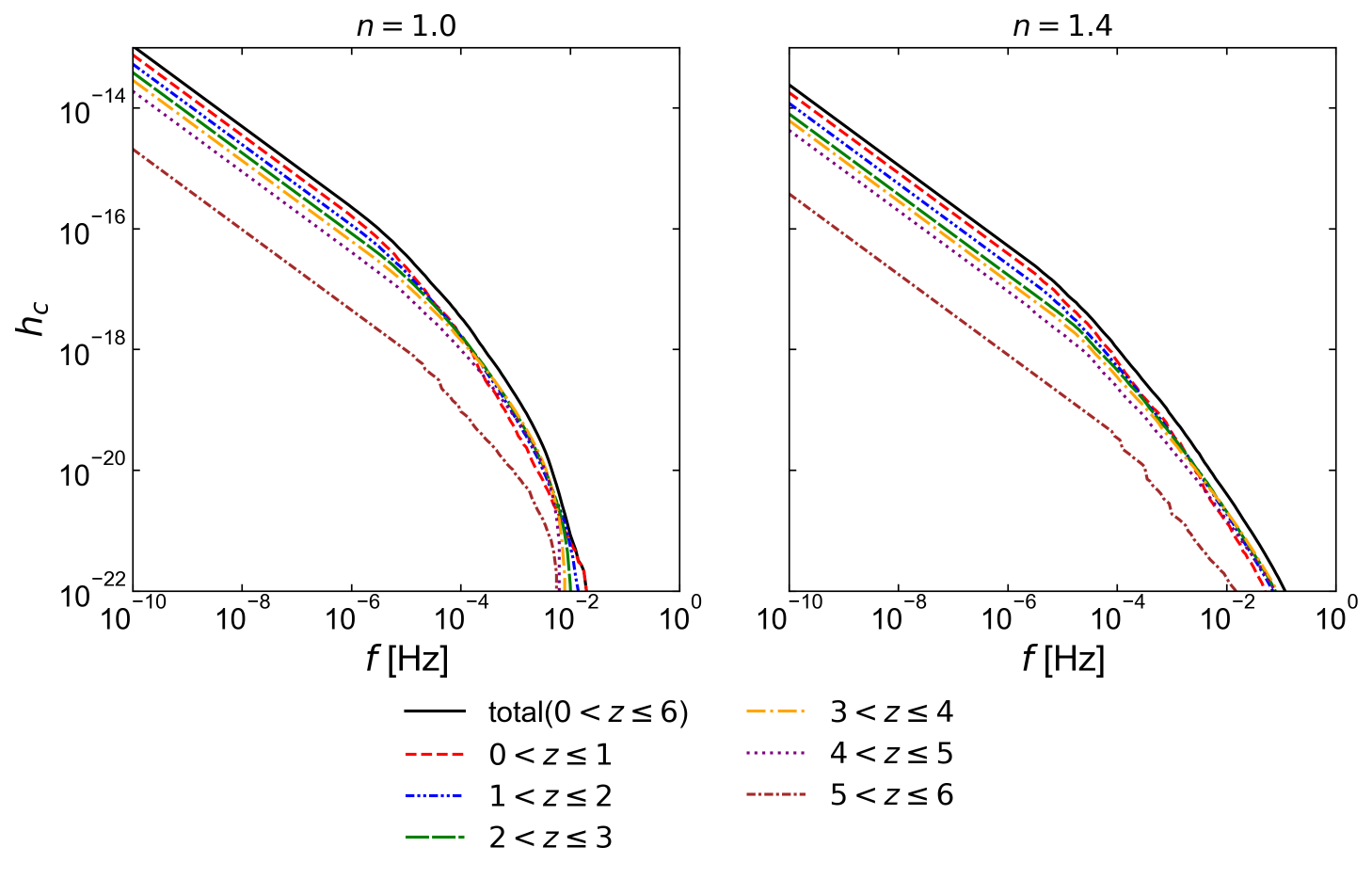}
  \caption{The characteristic strain amplitude of the SGWB from SMBH binaries for the cases with $n=1.0$ (left) and $1.4$ (right). $M_{\mathrm{lim}}=1.0\times 10^8 M_{\odot}h^{-1}$ is fixed. The upper and lower panels show the contribution from each mass bin and each redshift bin, respectively.}
  \label{fig:hc_108}
\end{figure}

Figure~\ref{fig:hc_108} shows the characteristic strain amplitudes for the cases with $n=1.0$ (left) and $1.4$ (right), respectively. The upper panels show the contribution from each mass bin in the amplitude, and the lower ones show the contribution from each redshift bin in the amplitude. In each panel, the total amplitude is shown by a black line. From this figure, one can find that the contribution from the coalescing SMBHs with masses in the range of $10^7 - 10^9 M_{\odot}$ at $0 \leq z \leq 1$ is dominant. This is simply because the larger system can emit the larger amplitude of GWs, as shown by Equation~\eqref{spec_inspiral} (see also, e.g. \citealt{2004ApJ...615...19E}, \citealt{2008MNRAS.390..192S}).
While there is a contribution from
the SMBH binaries with $M_{\rm BH} \gtrsim 10^9 M_\odot$ 
in the case with $n=1.0$, it is small. This is due to the small number of such heavier SMBHs, which can be seen in Figure~\ref{fig:HaloBHz0}. As for the dependence on $M_{\mathrm{lim}}$, as shown in Figure~\ref{fig:HaloBHz0}
the population of the SMBHs at the lower redshift is almost independent of $M_{\mathrm{lim}}$, and Figure~\ref{fig:hc_108} shows that the dominant contribution in the SGWB is coming from the coalescing SMBHs at the redshifts $0 \leq z \leq 1$. Thus, from these facts, the amplitude of the SGWB from the coalescing SMBHs in the frequency band of the PTA experiments ($f\sim 10^{-9} - 10^{-8}$ Hz) is almost independent of $M_{\mathrm{lim}}$.

\begin{figure}
\plottwo{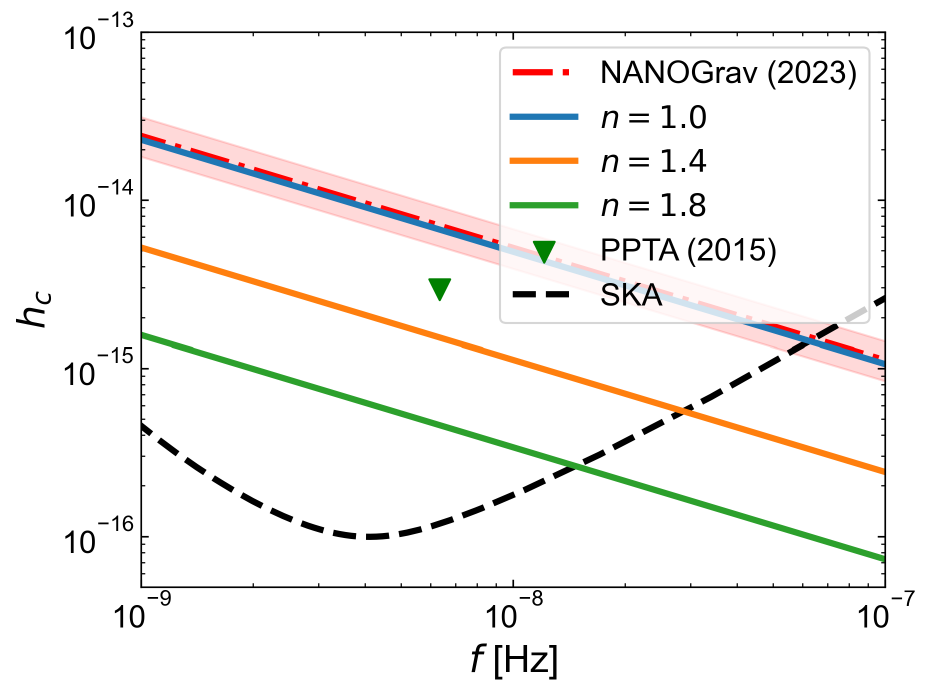}{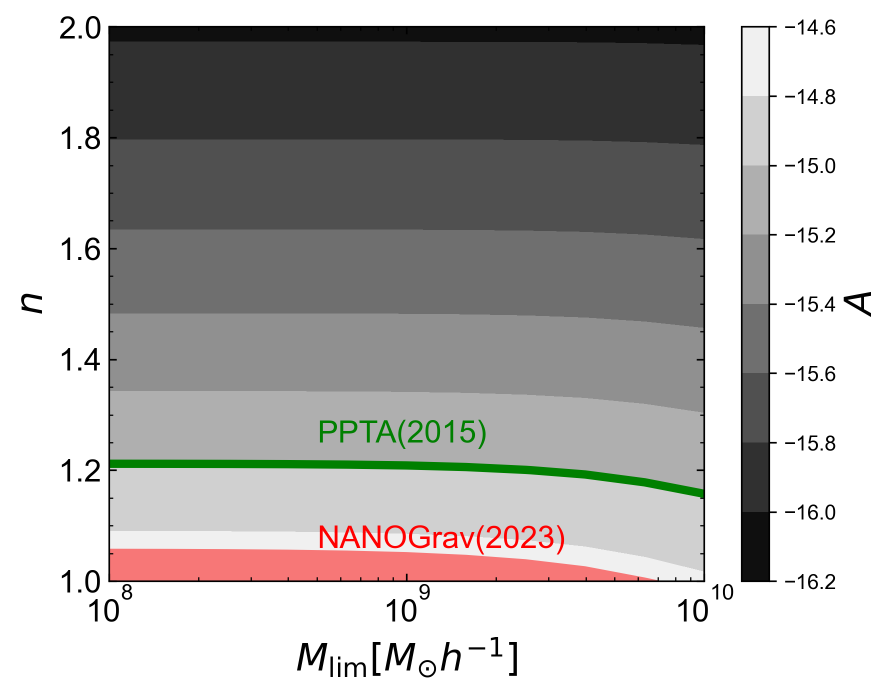}
\caption{
Comparison of the theoretical predictions in our coevolution model
and the PTA observational results.
The left panel shows the theoretical predictions
of the characteristic strain in the cases with $n=1.0$ (blue), $1.4$ (orange), and $1.8$ (green) for $M_{\mathrm{lim}} = 1.0 \times 10^8 M_\odot h^{-1}$, and the upper bound obtained from the PPTA project~(\citealt{2015Sci...349.1522S}) is represented by a triangle. We also include the possible signal of SGWB
corresponding to the common-spectrum process reported
from the recent NANOGrav experiment~(\citealt{2023ApJ...951L...8A})
represented by the red band and the sensitivity curve
expected for the future PTA experiment by SKA with the black dashed-dotted line (see, e.g., \citealt{2021JCAP...01..012C}).
The right panel shows the contours of the amplitude of the characteristic strain at $f=1 {\rm yr}^{-1}$ on the $n$-$M_{\mathrm{lim}}$ plane.
}
\label{fig:hc}
\end{figure}

In Figure~\ref{fig:hc}, we show the comparison of the theoretical predictions in our coevolution model and the PTA observational results.
The left panel shows the theoretical predictions
of the characteristic strain in the cases with $n=1.0$ (blue), $1.4$ (orange), and $1.8$ (green) for $M_{\mathrm{lim}} = 1.0 \times 10^8 M_\odot h^{-1}$, and the upper bound obtained from the PPTA project~(\citealt{2015Sci...349.1522S}) is represented by a triangle. 
We also include the possible signal of SGWB
corresponding to the common-spectrum process reported
from the NANOGrav experiment~(\citealt{2023ApJ...951L...8A})
represented by the red band and the sensitivity curve
expected for the future PTA experiment by SKA with the black dashed-dotted line (see, e.g., \citealt{2021JCAP...01..012C}).

The right panel shows the contours of the amplitude of the characteristic strain at $f=1 {\rm yr}^{-1}$ on the $n$-$M_{\mathrm{lim}}$ plane.
First, from the left panel, one can find that
the amplitude of SGWB becomes smaller as $n$ increases, which can be also
seen in Figure~\ref{fig:hc_108}.
This is simply because the typical mass of SMBHs in the lower redshift
becomes smaller for larger $n$ as shown in Figure~\ref{fig:HaloBHz0}.
As the result reported by the NANOGrav experiment, we simply employ the result assuming the simple power-law model as
\begin{equation}
h_c(f) = A \left( \frac{f}{1~{\rm yr}^{-1}} \right)^{-2/3},
\end{equation}
with $A=2.4_{-0.6}^{+0.7}\times 10^{-15}$ \citep{2023ApJ...951L...8A}.

The right panel in Figure~\ref{fig:hc} tells us that, in this model,  the current observational result obtained from NANOGrav (2023) implies $n \simeq 1.0$ and this implication is almost independent of the value of $M_{\mathrm{lim}}$. This means that the PTA experiments can provide information about the mass relation between the DM halos and the SMBHs at $z=6$. However, to keep the consistency with the local mass relation as mentioned in Section~\ref{sec:DM halo-SMBH}, we need to consider the gas accretion process. We will discuss this issue later in Section~\ref{subsec:caveat}.

\subsection{GW burst}

Let us discuss the detectability of the GW bursts from the individual SMBH mergers in our coevolution model
by using future planned GW experiments such as LISA. After the inspiraling phase discussed above, the SMBH binaries finally enter the merging phase and emit strong burst-like GWs. The waveform
of GWs emitted during this phase is difficult to estimate analytically, and is, in general, estimated by
calculations based on numerical relativity. Here, we employ IMRPhenomD \citep{2016PhRvD..93d4007K} using PyCBC\footnote{https://pycbc.org/} to generate the Fourier waveform of a burst.

The criteria for whether a GW burst is detected is closely related to the signal-to-noise ratio (SNR)
which is defined as (\citealt{2021arXiv210801167B})
\begin{equation}
\rho^2 \equiv \langle \mathrm{SNR}^2 \rangle = \mathrm{Re}\Big(4\int_{f_{\mathrm{min}}}^{f_{\mathrm{max}}}\frac{\frac{4}{5}A^2(f)}{S_{\mathrm{SciRD}}(f)}\dd f\Big),
\end{equation}
where $A$ is the amplitude of the burst in Fourier space and 
$S_{\mathrm{SciRD}}(f)$ is the official required sensitivity of LISA introduced in the LISA Science Requirements Document, given as (\citealt{2021arXiv210801167B})
\begin{equation}
\label{LISA}
\begin{split}
S_{\mathrm{SciRD}}(f) &= \frac{1}{2}\frac{20}{3}\Big(\frac{S_1(f)}{(2\pi f)^4}+S_2\Big)R(f),\\
&S_1(f) = 5.76\times10^{-48}\Big(1+\Big(\frac{f_1}{f}\Big)^2\Big)\mathrm{s}^{-4}\mathrm{Hz}^{-1},\\
&S_2(f) = 3.6\times 10^{-41}\mathrm{Hz}^{-1},\\
&R(f) = 1+\Big(\frac{f}{f_2}\Big)^2,\\
&f_1=0.4\mathrm{mHz},f_2 = 25\mathrm{mHz}.
\end{split}
\end{equation}
We adopt the detection threshold as $\rho_\mathrm{th} = 5$.

Thus, the number of detectable GW bursts per year coming from $z\sim z+dz$ can be calculated as
\begin{equation}
\label{BurstGW_Num}
\frac{\dd N_{\mathrm{mgr}}}{\dd z}dz = 4\pi d^2(z) c\dd z \int \dd M_1\dd M_2 \frac{\dd^3n_c}{\dd z\dd M_1\dd M_2}\theta(\rho-\rho_\mathrm{th}),
\end{equation}
where $d(z)$ is the comoving distance to the redshift $z$
and $\theta(\rho-\rho_\mathrm{th})$ is the step function to select the detectable signal with $\rho\geq \rho_\mathrm{th}$.

\begin{figure}[t!]
\plotone{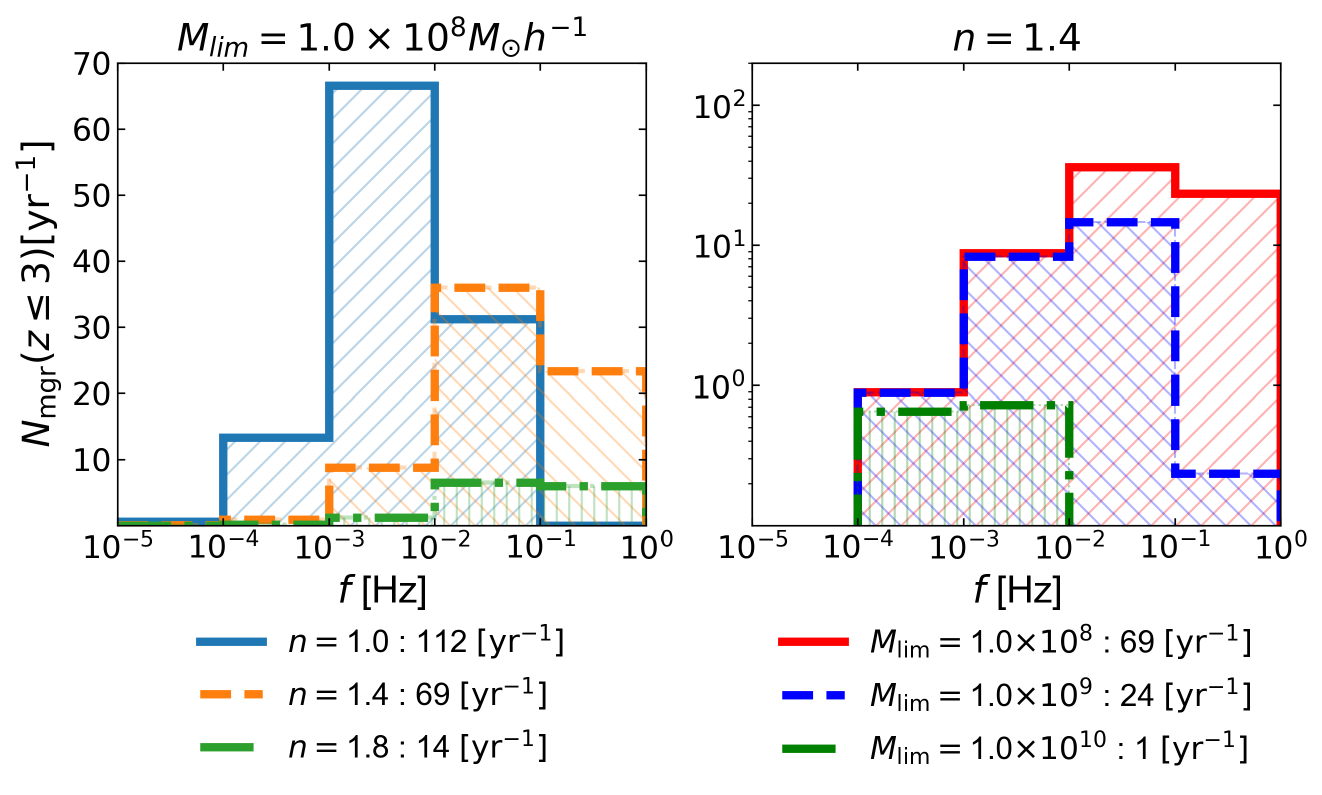}
\caption{The distribution of the expected number of the burst events which would be detected in the LISA 
with respect to the peak frequency.
The left panel shows the dependence on the power-law index in our coevolution model with fixing $M_{\mathrm{lim}} = 1.0 \times 10^8 M_\odot h^{-1}$.
The right one shows the result with changing the limiting mass for the case with $n=1.4$.
}
\label{fig:BurstGWf}
\end{figure}

Figure~\ref{fig:BurstGWf} shows the distribution of the expected number of the GW bursts that would be detected in the LISA with respect to the peak frequency. 
The left panel shows the result with changing the power-law index as $n=1.0$ (blue), $1.4$ (orange), and
$1.8$ (green). 
In the right panel, we change the limiting mass, $M_{\mathrm{lim}}$ from $1.0 \times 10^8 M_\odot h^{-1}$ to $1.0 \times 10^{10} M_\odot h^{-1}$.
As we have discussed in Section~\ref{sec:DM halo-SMBH},
in our DM-halo and SMBH coevolution model,
we assume the power-law relation given by Equation~\ref{init_powerlaw} at $z=6$
as the initial condition,
and then our calculation can not trace the halo merger history and the evolutionary history of the SMBHs before $z=6$.
Based on our criteria for the coalescing SMBHs discussed in the previous section, the coalescing SMBH binaries
associated with the mergers of halos which occur at $z \geq 6$ are expected to merge at $z \geq 3$.
Thus, to obtain a conservative estimation, here, we only count the burst events which occur at $z \leq 3$.
In our coevolution model, as can be seen in Figure~\ref{fig:HaloBHz6_108},
as $n$ increases
small halos host the SMBHs with smaller mass. In general, smaller halos exist in larger numbers,
and then for the larger $n$, the typical mass of the SMBHs in our Universe becomes smaller.
Since the peak frequency of the burst is inversely proportional to the mass of the binary system, which is roughly
given by
\begin{equation}
\label{peak_freq}
f_{\mathrm{peak}} \simeq 1.9 \times 10^{-4}\mathrm{Hz}\Big(\frac{1}{1+z}\Big)\Big(\frac{10^8M_{\odot}}{M_{\mathrm{tot}}}\Big).
\end{equation}
the distribution of the burst events would shift toward the higher frequency as $n$ increases
and eventually goes outside the LISA band.
As can be seen in the left panel of Figure~\ref{fig:BurstGWf}, 
therefore, the total number of detectable events decreases as $n$ increases.
As for the dependence on the limiting mass shown in the right panel, 
unlike the case of SGWB, the expected number of events
also depends on the limiting mass.
Basically, the increase in the limiting mass means that the number of halos hosting the SMBHs decreases,
and then the number of burst events should also decrease.

\begin{figure}[t!]
\centering
\includegraphics[width=0.6\textwidth]{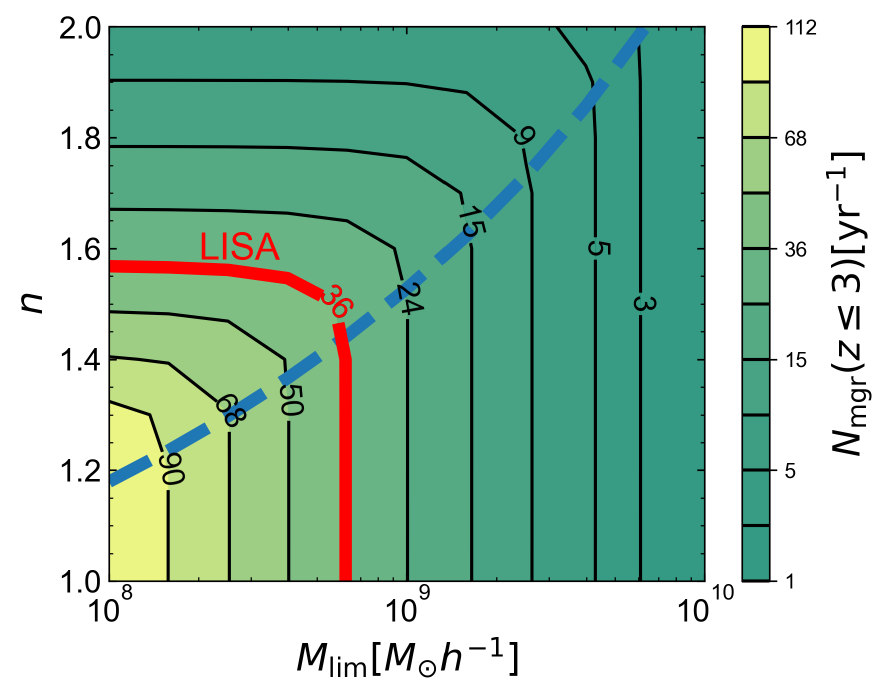}
\caption{The contours of the expected number of burst events in the LISA experiment
in the $n$-$M_{\mathrm{lim}}$ plane.
The red line denotes the example of future prospect of burst events per year detected by LISA. 
The blue dashed line is the boundary that divides the $n$-$M_\mathrm{lim}$ plane into two regions where the minimum SMBH mass at $z=6$ is lighter than $10^4M_\odot$ (upper side) or not (lower side).}
\label{fig:BurstGW}
\end{figure}

Figure~\ref{fig:BurstGW} is the contour plot for the expected number of burst events in the LISA experiment
in the $n$-$M_{\mathrm{lim}}$ plane. 
The figure shows that the dependence of the 
expected number of burst events is different between the upper and lower regions divided by the blue dashed line.
In the upper region, small dark matter halos whose masses are comparable to $M_{\rm lim}$ have SMBHs with a smaller mass than $10^4~M_\odot$ at $z=6$.
The LISA observation opens the sensitivity in the frequencies of the GW bursts from SMBH merger with masses larger than $10^4~M_\odot$. Therefore, although the mergers of small DM halos occupy a large part of the merger events, they produce GW bursts which is not in the LISA observation frequency. Since the steeper initial mass relation of the SMBH-DM relation~(larger power-law index $n$) provides
smaller SMBHs for small DM halos,
the expected number of burst events in the LISA experiment becomes
small with the increase of $n$ value.
In the lower region of Figure~\ref{fig:BurstGW},
small DM halos whose masses are comparable to $M_{\rm lim}$ can host SMBHs with larger mass than $10^4~M_\odot$ at $z=6$.
Therefore, the merger of small DM halos can produce the GW burst in
the LISA frequency range
and the expected number does not depend on $n$.
However, as $M_{\rm lim}$ increases, the merger rate between DM halos that host SMBH 
decreases. Therefore,
larger $M_{\rm lim}$ provides a small expected number of the GW burst events in the LISA experiment in the lower region.

Basically, from this figure, once the number of burst events from $z\leq 3$
is specified (as shown in the redline in Figure~\ref{fig:BurstGW}), we can obtain the constraint on the mass relation between the
DM halos and SMBHs parameterized by $n$ and $M_{\mathrm{lim}}$. 
However, note that in our model, we only consider the mergers for the mass evolution of SMBHs, neglecting the gas accretion onto the BHs. In Section~\ref{subsec:caveat}, taking into account the actual mass growth of SMBH, we discuss the approach of parameter constraint from this result of event rate for LISA.

\subsection{Gas accretion effect}\label{subsec:caveat}
In our model, we include only SMBH mergers for the mass growth process of SMBHs, as mentioned in the last part of Section~\ref{sec:DM halo-SMBH}.
However, the gas accretion is also important for the SMBH mass growth.
In this section, we discuss the impact of the gas accretion on our results. 

First, we consider the effect of the gas accretion on the SGWB with PTAs.
In our model, the amplitude of the SGWB mainly depends on the event rate of the massive SMBH mergers in the mass range with $10^7 M_\odot \leq M_\mathrm{BH,tot}$ and $0\leq z \leq 2$ as shown in Figure \ref{fig:hc_108}.
Including the gas accretion in our model would lead to the mass enhancement of all SMBHs
and increase the event rate of massive SMBH mergers.
Accordingly, the gas accretion may result in the enhancement of the GW background amplitude.
In that sense, we expect that our model that does not include the gas accretion predicts the lower limit of SGWB
due to the SMBH mergers
and the red band in Figure~\ref{fig:hc} shifts upward, depending on the gas accretion efficiency. Therefore, from NANOGrav (2023), we can constrain the lower limit on $n$ as $n\gtrsim 1.0$.

Next, we discuss the detectability of GW bursts by considering gas accretion.
LISA can detect the GW burst from the SMBH merger with a mass larger than $10^4 M_\odot$.
In the region above the blue dashed line in Figure~\ref{fig:BurstGW},
small DM halos host SMBHs with mass smaller than $10^4 M_\odot$. 
The gas accretion would enhance the mass growth of SMBHs.
As a result, the number of small DM halos having SMBH with a mass larger than $10^4 M_\odot$ can increase,
and the event rate of the GW burst detectable by LISA also becomes large.
Therefore, in the upper region, the gas accretion can be expected to increase the number of the GW burst events for LISA, depending on the gas accretion efficiency in the SMBH mass growth.
On the other hand, in the lower region, 
small DM halos already host SMBHs with a mass larger than $10^4 M_\odot$ at $z=6$.
Thus, although gas accretion may increase the mass of SMBH, there would be little difference in the number of GW burst events detectable by LISA with or without gas accretion.
In that sense,
in the lower region of  Figure~\ref{fig:BurstGW},
the gas accretion would not affect the expected number of GW burst events for LISA.

\section{summary and discussion} \label{sec:Conclusion}

In this paper, we have investigated the potential of GW observations 
to probe the mass relation between DM halos and SMBHs.
The mass relation in high redshifts could be a key to revealing the DM halo-SMBH coevolution and
the physics of the SMBH mass growth.

We have constructed the DM halo-SMBH coevolution 
model according to the merger tree based on the extended Press-Schechter formalism. In our model, the SMBH mass grows
through the SMBH coalescence triggered by the DM halo mergers.
The initial mass relation of SMBHs and DM halos is
given as a power-law type relation with
two model parameters;
one is the power-law index of the relation at $z=6$, $n$, and
the other is the lowest mass of DM halos that can host SMBHs, $M_{\rm lim}$.
Our model can roughly connect the mass relation
observed in high redshift~($z \sim 6$) to that
in the present epoch.

To account for the dynamical friction to SMBHs,
we consider the time delay of the SMBH coalescence after the DM halo mergers in our model.
This method allows us to treat multiple SMBHs and coalescences in a single DM halo, although the binary merger tree method is used for the DM halo mergers.

Using our SMBH-DM halo coevolution model, we 
have evaluated two types of GW signals; the GW background in the $\sim {\rm nHz}$ frequency range and the GW burst.
The former can be probed by PTA experiments
and the latter can be detected by the future space gravitational observatory, LISA.

We have shown that the amplitude of the GW background strongly depends on the power-law index of the initial mass relation, $n$, while it is almost immune to $M_{\rm lim}$.
We found that the observational result obtained from NANOGrav (2023) implies $n \simeq 1.0$ and it means that small DM halos at high redshift have relatively massive BHs in our model.
We have also demonstrated that near future PTA experiments such as SKA can provide a stronger constraint on $n$.

For the GW bursts from SMBH coalescences, we have calculated the number of 
detectable bursts per year with LISA sensitivity.
Our results show that the number of detectable GW bursts depends on both $n$ and $M_{\rm lim}$. Therefore, once we obtain an upper limit or constraint on the number of GW bursts per year, we can impose the limits on 
$n$ and $M_{\rm lim}$ from the LISA observation, which can then be used to constrain the formation and evolution processes of SMBHs at high redshifts.

Future galaxy and QSO surveys can provide constraints on the mass relation between SMBHs and DM halos
in high redshifts. 
However, in comparison with these observations, GW observations can constrain the mass relation even for the smaller mass region of DM halos and SMBHs that galaxy and QSO surveys can not probe.
On the other hand, it is important to construct a precise coevolution model of DM halos and SMBHs
to obtain the robust constraint on the mass relation in high redshifts.

In our coevolution model, we do not take into account the gas accretion on SMBHs, which is one of the important processes for SMBH mass growth. We can expect that for a fixed power-law index $n$ the amplitude of SGWB in the PTA frequency band would be enhanced due to the gas accretion, and the favorable value of $n$ becomes large depending on the accretion efficiency. Therefore, we can put the lower bound on $n$ as $n\gtrsim 1.0$. As for the LISA experiment, the dependence of the expected number of GW bursts on the model parameters, $n$ and $M_{\rm lim}$ would be changed by the gas accretion effect. Considering the behavior of the total event of GW bursts under this process, we can find the specific parameter regions ruled out from the future LISA experiment. Basically, in order to obtain more realistic constraints on the parameters $n$ and $M_\mathrm{lim}$, we need to consider the effect of the gas accretion process carefully.
Although it is difficult to model the gas accretion in high redshifts,
the detailed observation of the luminosity function of distant QSOs can provide us the 
information on the efficiency of the gas accretion in high redshifts (see e.g., \citealt{2020MNRAS.498.5652K}).
We will address these issues in our future work.

In this work, we do not take into account the detailed 
parameters of the SMBH merger, e.g., the spins and eccentricity of BH binaries.
The spins of the BH binary are related to the final BH recoil.
The strong final BH recoil in the SMBH coalescence can eject one of the two SMBHs
from the host DM halo.
The eccentricity of the BH binary affects the 
spectrum of the GW radiation. Larger eccentricity leads to a decrease 
in the amplitude of the GW radiation in the PTA frequency range~\citep{2007PThPh.117..241E, 2014MNRAS.442...56R}.
Therefore, such parameters of the SMBH merger affect the amplitude of the GW background or
the event rate of the GW bursts.
Also, the orbital evolution via the three-body interaction between the SMBH binary and stars in the system affects the spectrum in the low-frequency \citep{2004ApJ...611..623S}.
Recent PTA observations such as NANOGrav(2023) (see e.g. \citealt{2023ApJ...951L...8A} and \citealt{2023ApJ...952L..37A}) imply the importance of the effect of eccentricity and the orbital evolution via environmental interaction from the view of the spectrum of observed GWB. Furthermore, if triple BH interaction occurs in a system, one of three BHs will be ejected, and then the radius of the other two BHs will shrink and lead to their coalescence \citep{2003ApJ...582..559V, 2018MNRAS.477.2599B}. 
Thus, triple BH interactions also have a possibility to affect the SMBH merger rate. As future work, we should consider how these affect SMBH evolution and future GW observations.

\section*{Acknowledgements}
We thank K. Shimasaku for providing us with the compiled data of the halo-SMBH mass relation. We also thank E. Komatsu and D. Toyouchi for meaningful discussions.
This work is supported in part by the JSPS grant numbers
20K03968, 21K03533 and 21H04467,  Core-to-Core Program grant numbers JPJSCCA20200002, JPJSCCA20200003,  and JST FOREST Program JPMJFR20352935.

\appendix

\section{Dynamical Friction Timescale between Black Hole and its Host Galaxy}\label{sec:App.A}

In our coevolution model discussed in Section~\ref{sec:DM halo-SMBH}, we employed the model
where the timescale of SMBH mergers is given by
that of dynamical friction between host DM halos given by Equation~\eqref{tdf}, following \cite{2003ApJ...582..559V}. 

Recently, in \cite{2023MNRAS.523..758C} that investigated the merger rate by using cosmological hydrodynamical simulation, they adopted the dynamical friction timescale between a lighter BH and the stellar component in the host galaxy 
which is assumed to have the heavier BH at the center,
which is given by
\begin{equation}
t_{\mathrm{df}} = 0.67\mathrm{Gyr}\Big(\frac{a}{4\mathrm{kpc}}\Big)^2\Big(\frac{\sigma_\ast}{100\mathrm{km/s}}\Big)\Big(\frac{10^8M_\odot}{M_{\mathrm{BH,L}}}\Big)\frac{1}{\Lambda},
\label{new_tdf}
\end{equation}
where $a$ is the separation of two coalescing BHs, $M_{\ast}$ is the total stellar mass of the host galaxy, $\sigma_\ast = (0.25GM_{\ast}/R_{\mathrm{eff}})^{1/2}$ is the velocity dispersion, $R_{\mathrm{eff}} = 0.1 r_{\mathrm{vir}}$, $M_{\mathrm{BH,L}}$ is the mass of the lighter BH, and $\Lambda=\ln(1+M_{\ast}/M_{\mathrm{BH,L}})$.

\begin{figure}
\plottwo{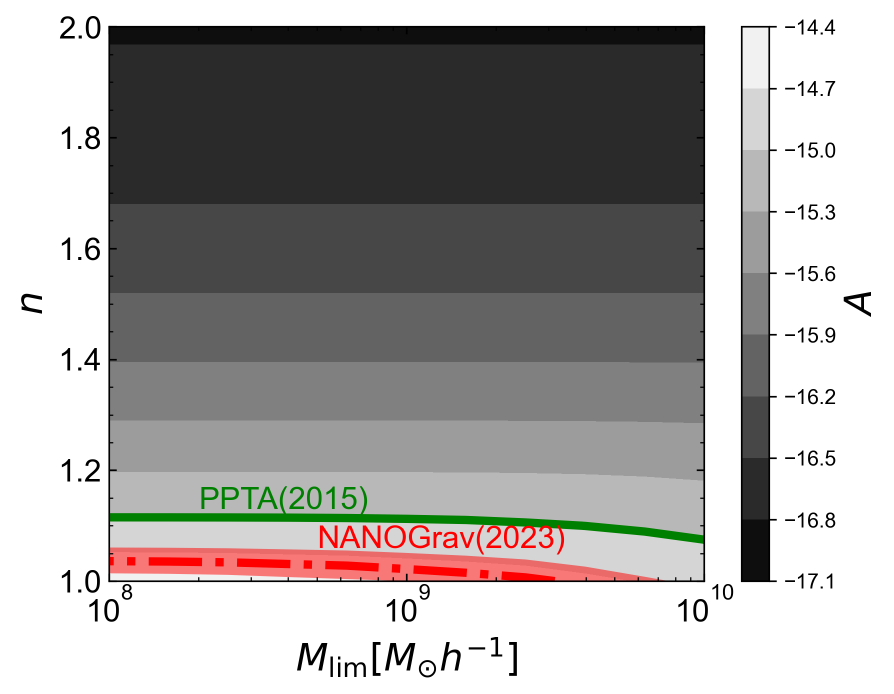}{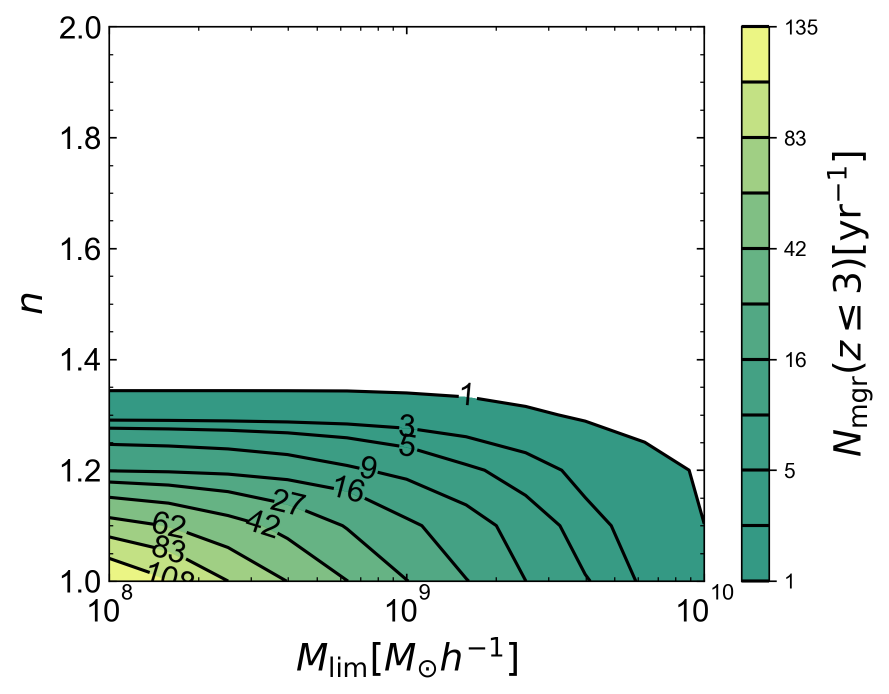}
\caption{Theoretical predictions of SGWB $A$ (left panel) and the event rate of GW bursts $N_{\rm mgr}$ (right panel) for the different dynamical friction timescale given by Equation~\eqref{new_tdf}.}
\label{fig:app_GW}
\end{figure}

Employing this timescale for the SMBH mergers, 
in particular, 
the merger timescale for lighter BHs would be longer,  
compared with the timescale given by Equation~\eqref{tdf},
and some binaries with small masses could not merge within cosmic time.
Thus, the resultant GWs in the coevolution model can be relatively sensitive to the masses of SMBHs
and also the power-law index $n$.
The results with the timescale given by Equation~\eqref{new_tdf} are shown in Figure~\ref{fig:app_GW}.
Here, we 
assume that $M_{\ast} = (\Omega_{\mathrm{b}}/\Omega_{m})M_\mathrm{halo}$, where $M_\mathrm{halo} = M_\mathrm{H}+M_\mathrm{L}$ is the total halo mass of the new system (see Figure~\ref{fig:DMhalo-SMBH Merger}), and $a=R_{\mathrm{eff}}$ for simplicity.
As well as our model, we neglect the SMBH merger whose dynamical friction timescale exceeds the Hubble time at that time.

In this figure, the left panel shows the contours of the amplitude of SGWB in $n$-$M_{\mathrm{lim}}$ plane. 
The amplitude of the SGWB is dominated by the contribution from the large mass BHs, and no matter which timescale is adopted, they always coalesce within the timescale considered, and therefore, the results are not significantly affected. 
The right panel of Figure~\ref{fig:app_GW} shows the event rate of GW bursts for each $n$ and $M_{\mathrm{lim}}$.
For the GW bursts, the contribution from small mass BHs dominates, and hence the change in timescale seems to have an impact, especially when $n$ is large and small mass BHs are more abundant. However, in fact, when $n$ is large, the expected number of GW bursts was small, and therefore, we find that the change in timescale does not have a large impact on the results. 

If taking the additional mass growth of SMBH into account, then the timescale of all SMBH mergers will become shorter. 
Therefore, naively our prediction of SGWB and the total event rate of GW bursts here is considered conservative.

\section{Hardening}\label{sec:App.B}

In our model, the delay time of SMBH merger from the merger of the host DM halos is given by the dynamical friction timescale calculated by Equation~\eqref{tdf}. However, 
there exists the "hardening phase"~(see e.g.,~\citealt{2014SSRv..183..189C}) between the phase where the dynamical friction leads to the shrink of the binary separation and the inspiral GW phase.
In the hardening phase, 
the binary separation shrinks through 
the three-body interactions between the SMBH binary and the surrounding stars.
\cite{2019MNRAS.487.4985B} have investigated the shaking timescale in the hardening phase at the local universe.
They have shown that the timescale is comparable with or exceeds the Hubble time at $z=0$ in the case of the SMBH binary with $M_\mathrm{BH,tot}\lesssim 10^6 M_\odot$.
Therefore, the hardening phase might affect the event rate for GW emissions.
In this section, we discuss the impact of the hardening phase on our results through the delay time of SMBH mergers from the merger of the host DM halos. 

The hardening timescale, which is the timescale of the binary separation shrinking in the hardening phase, 
is given by the timescale of this three-body interaction process.
In \cite{2019MNRAS.487.4985B}, they have investigated the hardening timescales at the local universe, assuming the different stellar density profiles in a galaxy~(see Figure~5 in their work). Their results suggest that the hardening timescale depends on the mass of a BH binary.

\begin{figure}[t!]
\centering
\includegraphics[width=0.6\textwidth]{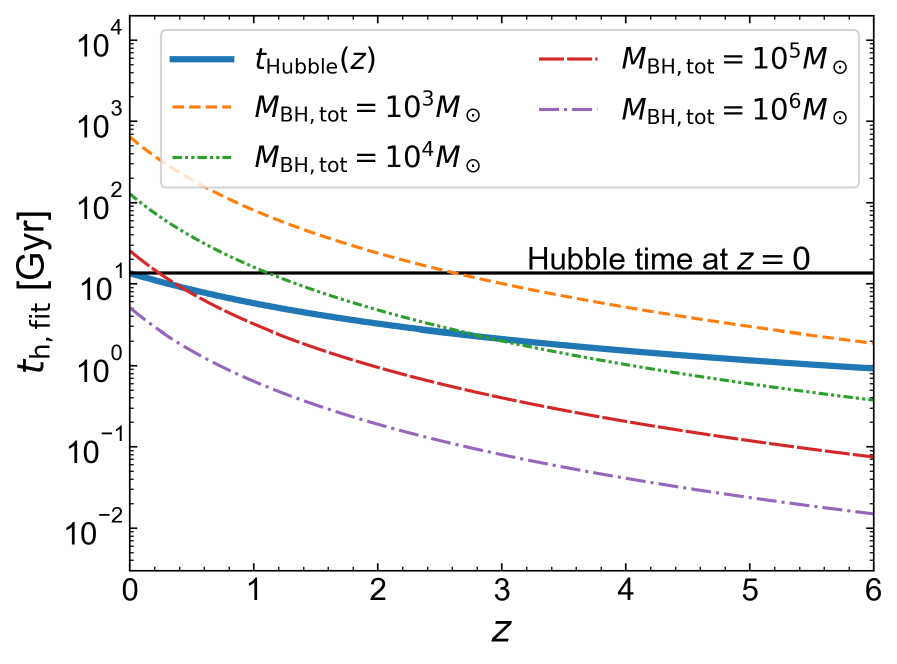}
\caption{Hardening timescales with different $M_\mathrm{BH,tot}$ as a function of redshift. 
From top to bottom,
the hardening timescales are for $M_\mathrm{BH,tot}=10^3M_\odot$, $10^4M_\odot$, $10^5M_\odot$, and $10^6M_\odot$, respectively. 
For comparison, the thick solid line (blue) shows the Hubble time as a function of $z$ and the thin solid line (black) corresponds to the Hubble time at $z=0$.
}
\label{fig:th_fit}
\end{figure}

To adopt their result in our model,
we construct the fitting formula to the results in Figure 5 of ~\cite{2019MNRAS.487.4985B} as
\begin{equation}
t_\mathrm{h} = 5.1\mathrm{Gyr}\Big(\frac{M_\mathrm{BH,tot}}{10^6M_\odot}\Big)^{-0.7}(1+z)^{-3},
\label{th_fit}
\end{equation}
where we include the redshift dependence suggested in~\cite{2017JPhCS.840a2025M}.
We plot the dependence of the hardening timescale on $z$ and $M_\mathrm{BH,tot}$ in Figure~\ref{fig:th_fit}.

The hardening timescale delays the SMBH merger and the GW emission from the host DM halo merger.
In order to investigate the impact of the hardening timescale,
we calculate the SGWB and the GW bursts in the following way. 
In the merger tree, when we find that a DM halo merger at $t=t_\mathrm{mg}$, we calculate the dynamical friction timescale $t_\mathrm{df}$ at $t=t_\mathrm{mg}$ as described in 
Section~\ref{subsec:Our_Model}.
Then we calculate the hardening timescale $t_\mathrm{h}$  at $t=t_\mathrm{mg}+t_\mathrm{df}$.
If the calculated timescale $t_\mathrm{h}$ is smaller than the Hubble time at that time, 
we take into account the coalescence of the SMBH binary at $t = t_\mathrm{mg} + t_\mathrm{df} + t_\mathrm{h}$ and its GW radiation.
Otherwise, we neglect the SMBH merger.

\begin{figure}
\plottwo{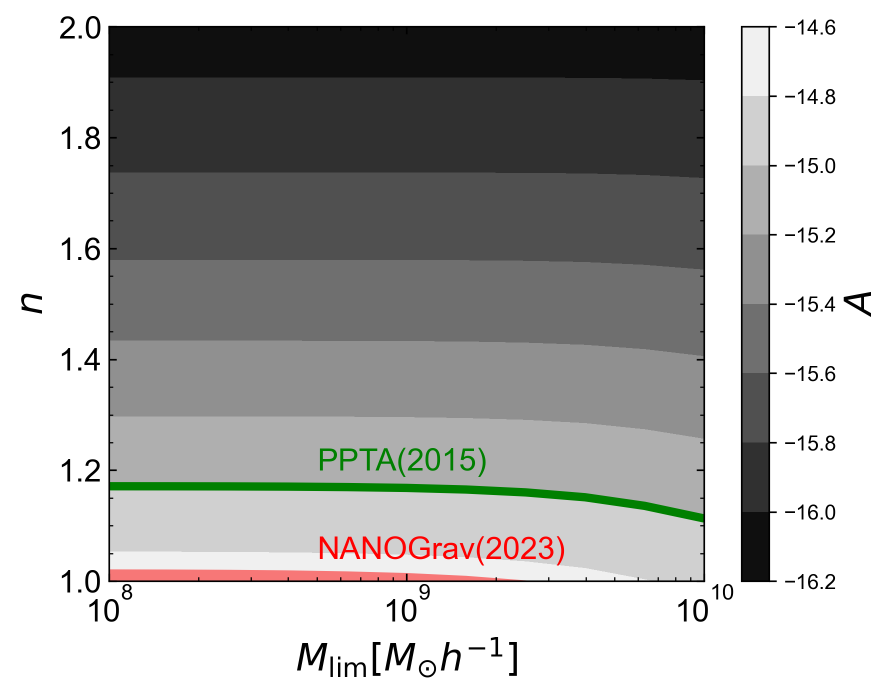}{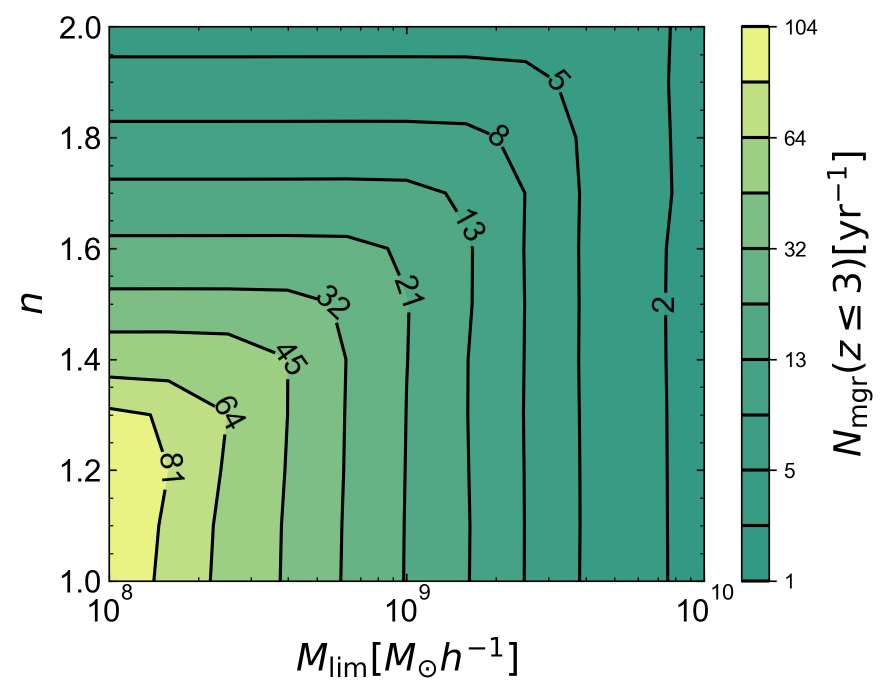}
\caption{Theoretical predictions of SGWB amplitude $A$ at $f=1 {\rm yr}^{-1}$ (left panel) and the event rate of GW bursts $N_{\rm mgr}$ (right panel) including the effect of the hardening phase.}
\label{fig:th_GW}
\end{figure}

The left side of Figure~\ref{fig:th_GW} shows the amplitude of the SGWB on the $n$-$M_\mathrm{lim}$ plane, including the hardening timescale.
Comparing with the right panel of Figure~\ref{fig:hc}, we find that the amplitude of the SGWB becomes slightly lower with the hardening timescale than without the hardening timescale. This is because the number of SMBH mergers with $M_\mathrm{BH,tot} \lesssim 10^6 M_\odot$ at $0\leq z \leq 2$ decreases due to the additional hardening timescale. 
The right panel of Figure~\ref{fig:th_GW} represents the expected number of GW burst events that LISA can detect on the 
$n$-$M_\mathrm{lim}$ plane. 
Compared with Figure~\ref{fig:BurstGW}, Figure~\ref{fig:th_GW} shows that the total event rate at $0\leq z \leq 3$ also slightly decreases at all regions on $n$-$M_\mathrm{lim}$ plane due to the hardening timescale. However, its dependency on $n$ and $M_\mathrm{lim}$ is almost the same as that without the hardening timescale in Figure~\ref{fig:BurstGW}. 
Although considering the hardening timescale delays the SMBH mergers and the GW emission, we can conclude that the effects are limited and do not affect
our results significantly.

In the last part of this section, it is worth mentioning the effect of the delay timescale on the DM halo mergers at $z\geq6$.
Our model can count only the SMBH merger event induced by DM halo mergers at $z\leq6$. However, due to the delay of the SMBH merger from DM halo mergers,
the mergers of DM halos at $z\geq6$ can trigger SMBH mergers, and the GW burst at $0\leq z \leq 3$.
Thus, compared to the observational results, our prediction here basically underestimates the amplitude of SGWB and the total event rate of GW bursts.
\bibliography{ref}

\begin{thebibliography}{}
\expandafter\ifx\csname natexlab\endcsname\relax\def\natexlab#1{#1}\fi
\providecommand{\url}[1]{\href{#1}{#1}}
\providecommand{\dodoi}[1]{doi:~\href{http://doi.org/#1}{\nolinkurl{#1}}}
\providecommand{\doeprint}[1]{\href{http://ascl.net/#1}{\nolinkurl{http://ascl.net/#1}}}
\providecommand{\doarXiv}[1]{\href{https://arxiv.org/abs/#1}{\nolinkurl{https://arxiv.org/abs/#1}}}

\bibitem[{{Agarwal} {et~al.}(2019){Agarwal}, {Cullen}, {Khochfar}, {Ceverino}, \& {Klessen}}]{2019MNRAS.488.3268A}
{Agarwal}, B., {Cullen}, F., {Khochfar}, S., {Ceverino}, D., \& {Klessen}, R.~S. 2019, \mnras, 488, 3268, \dodoi{10.1093/mnras/stz1347}

\bibitem[{{Agazie} {et~al.}(2023{\natexlab{a}}){Agazie}, {Anumarlapudi}, {Archibald}, {Arzoumanian}, {Baker}, {B{\'e}csy}, {Blecha}, {Brazier}, {Brook}, {Burke-Spolaor}, {Burnette}, {Case}, {Charisi}, {Chatterjee}, {Chatziioannou}, {Cheeseboro}, {Chen}, {Cohen}, {Cordes}, {Cornish}, {Crawford}, {Cromartie}, {Crowter}, {Cutler}, {Decesar}, {Degan}, {Demorest}, {Deng}, {Dolch}, {Drachler}, {Ellis}, {Ferrara}, {Fiore}, {Fonseca}, {Freedman}, {Garver-Daniels}, {Gentile}, {Gersbach}, {Glaser}, {Good}, {G{\"u}ltekin}, {Hazboun}, {Hourihane}, {Islo}, {Jennings}, {Johnson}, {Jones}, {Kaiser}, {Kaplan}, {Kelley}, {Kerr}, {Key}, {Klein}, {Laal}, {Lam}, {Lamb}, {Lazio}, {Lewandowska}, {Littenberg}, {Liu}, {Lommen}, {Lorimer}, {Luo}, {Lynch}, {Ma}, {Madison}, {Mattson}, {McEwen}, {McKee}, {McLaughlin}, {McMann}, {Meyers}, {Meyers}, {Mingarelli}, {Mitridate}, {Natarajan}, {Ng}, {Nice}, {Ocker}, {Olum}, {Pennucci}, {Perera}, {Petrov}, {Pol}, {Radovan}, {Ransom}, {Ray}, {Romano}, {Sardesai}, {Schmiedekamp}, {Schmiedekamp},
  {Schmitz}, {Schult}, {Shapiro-Albert}, {Siemens}, {Simon}, {Siwek}, {Stairs}, {Stinebring}, {Stovall}, {Sun}, {Susobhanan}, {Swiggum}, {Taylor}, {Taylor}, {Turner}, {Unal}, {Vallisneri}, {van Haasteren}, {Vigeland}, {Wahl}, {Wang}, {Witt}, {Young}, \& {Nanograv Collaboration}}]{2023ApJ...951L...8A}
{Agazie}, G., {Anumarlapudi}, A., {Archibald}, A.~M., {et~al.} 2023{\natexlab{a}}, \apjl, 951, L8, \dodoi{10.3847/2041-8213/acdac6}

\bibitem[{{Agazie} {et~al.}(2023{\natexlab{b}}){Agazie}, {Anumarlapudi}, {Archibald}, {Baker}, {B{\'e}csy}, {Blecha}, {Bonilla}, {Brazier}, {Brook}, {Burke-Spolaor}, {Burnette}, {Case}, {Casey-Clyde}, {Charisi}, {Chatterjee}, {Chatziioannou}, {Cheeseboro}, {Chen}, {Cohen}, {Cordes}, {Cornish}, {Crawford}, {Cromartie}, {Crowter}, {Cutler}, {D'Orazio}, {Decesar}, {Degan}, {Demorest}, {Deng}, {Dolch}, {Drachler}, {Ferrara}, {Fiore}, {Fonseca}, {Freedman}, {Gardiner}, {Garver-Daniels}, {Gentile}, {Gersbach}, {Glaser}, {Good}, {G{\"u}ltekin}, {Hazboun}, {Hourihane}, {Islo}, {Jennings}, {Johnson}, {Jones}, {Kaiser}, {Kaplan}, {Kelley}, {Kerr}, {Key}, {Laal}, {Lam}, {Lamb}, {Lazio}, {Lewandowska}, {Littenberg}, {Liu}, {Luo}, {Lynch}, {Ma}, {Madison}, {McEwen}, {McKee}, {McLaughlin}, {McMann}, {Meyers}, {Meyers}, {Mingarelli}, {Mitridate}, {Natarajan}, {Ng}, {Nice}, {Ocker}, {Olum}, {Pennucci}, {Perera}, {Petrov}, {Pol}, {Radovan}, {Ransom}, {Ray}, {Romano}, {Runnoe}, {Sardesai}, {Schmiedekamp}, {Schmiedekamp},
  {Schmitz}, {Schult}, {Shapiro-Albert}, {Siemens}, {Simon}, {Siwek}, {Stairs}, {Stinebring}, {Stovall}, {Sun}, {Susobhanan}, {Swiggum}, {Taylor}, {Taylor}, {Turner}, {Unal}, {Vallisneri}, {Vigeland}, {Wachter}, {Wahl}, {Wang}, {Witt}, {Wright}, {Young}, \& {Nanograv Collaboration}}]{2023ApJ...952L..37A}
---. 2023{\natexlab{b}}, \apjl, 952, L37, \dodoi{10.3847/2041-8213/ace18b}

\bibitem[{{Amaro-Seoane} {et~al.}(2017){Amaro-Seoane}, {Audley}, {Babak}, {Baker}, {Barausse}, {Bender}, {Berti}, {Binetruy}, {Born}, {Bortoluzzi}, {Camp}, {Caprini}, {Cardoso}, {Colpi}, {Conklin}, {Cornish}, {Cutler}, {Danzmann}, {Dolesi}, {Ferraioli}, {Ferroni}, {Fitzsimons}, {Gair}, {Gesa Bote}, {Giardini}, {Gibert}, {Grimani}, {Halloin}, {Heinzel}, {Hertog}, {Hewitson}, {Holley-Bockelmann}, {Hollington}, {Hueller}, {Inchauspe}, {Jetzer}, {Karnesis}, {Killow}, {Klein}, {Klipstein}, {Korsakova}, {Larson}, {Livas}, {Lloro}, {Man}, {Mance}, {Martino}, {Mateos}, {McKenzie}, {McWilliams}, {Miller}, {Mueller}, {Nardini}, {Nelemans}, {Nofrarias}, {Petiteau}, {Pivato}, {Plagnol}, {Porter}, {Reiche}, {Robertson}, {Robertson}, {Rossi}, {Russano}, {Schutz}, {Sesana}, {Shoemaker}, {Slutsky}, {Sopuerta}, {Sumner}, {Tamanini}, {Thorpe}, {Troebs}, {Vallisneri}, {Vecchio}, {Vetrugno}, {Vitale}, {Volonteri}, {Wanner}, {Ward}, {Wass}, {Weber}, {Ziemer}, \& {Zweifel}}]{2017arXiv170200786A}
{Amaro-Seoane}, P., {Audley}, H., {Babak}, S., {et~al.} 2017, arXiv e-prints, arXiv:1702.00786, \dodoi{10.48550/arXiv.1702.00786}

\bibitem[{{Antoniadis} {et~al.}(2023){Antoniadis}, {Arumugam}, {Arumugam}, {Babak}, {Bagchi}, {Bak Nielsen}, {Bassa}, {Bathula}, {Berthereau}, {Bonetti}, {Bortolas}, {Brook}, {Burgay}, {Caballero}, {Chalumeau}, {Champion}, {Chanlaridis}, {Chen}, {Cognard}, {Dandapat}, {Deb}, {Desai}, {Desvignes}, {Dhanda-Batra}, {Dwivedi}, {Falxa}, {Ferdman}, {Franchini}, {Gair}, {Goncharov}, {Gopakumar}, {Graikou}, {Grie{\ss}meier}, {Guillemot}, {Guo}, {Gupta}, {Hisano}, {Hu}, {Iraci}, {Izquierdo-Villalba}, {Jang}, {Jawor}, {Janssen}, {Jessner}, {Joshi}, {Kareem}, {Karuppusamy}, {Keane}, {Keith}, {Kharbanda}, {Kikunaga}, {Kolhe}, {Kramer}, {Krishnakumar}, {Lackeos}, {Lee}, {Liu}, {Liu}, {Lyne}, {McKee}, {Maan}, {Main}, {Mickaliger}, {Nitu}, {Nobleson}, {Paladi}, {Parthasarathy}, {Perera}, {Perrodin}, {Petiteau}, {Porayko}, {Possenti}, {Prabu}, {Quelquejay Leclere}, {Rana}, {Samajdar}, {Sanidas}, {Sesana}, {Shaifullah}, {Singha}, {Speri}, {Spiewak}, {Srivastava}, {Stappers}, {Surnis}, {Susarla}, {Susobhanan}, {Takahashi},
  {Tarafdar}, {Theureau}, {Tiburzi}, {van der Wateren}, {Vecchio}, {Venkatraman Krishnan}, {Verbiest}, {Wang}, {Wang}, \& {Wu}}]{2023arXiv230616214A}
{Antoniadis}, J., {Arumugam}, P., {Arumugam}, S., {et~al.} 2023, arXiv e-prints, arXiv:2306.16214, \dodoi{10.48550/arXiv.2306.16214}

\bibitem[{{Babak} {et~al.}(2021){Babak}, {Hewitson}, \& {Petiteau}}]{2021arXiv210801167B}
{Babak}, S., {Hewitson}, M., \& {Petiteau}, A. 2021, arXiv e-prints, arXiv:2108.01167, \dodoi{10.48550/arXiv.2108.01167}

\bibitem[{{Bansal} {et~al.}(2023){Bansal}, {Ichiki}, {Tashiro}, \& {Matsuoka}}]{2023MNRAS.523.3840B}
{Bansal}, A., {Ichiki}, K., {Tashiro}, H., \& {Matsuoka}, Y. 2023, \mnras, 523, 3840, \dodoi{10.1093/mnras/stad1608}

\bibitem[{{Barkana} \& {Loeb}(2001)}]{2001PhR...349..125B}
{Barkana}, R., \& {Loeb}, A. 2001, \physrep, 349, 125, \dodoi{10.1016/S0370-1573(01)00019-9}

\bibitem[{{Biava} {et~al.}(2019){Biava}, {Colpi}, {Capelo}, {Bonetti}, {Volonteri}, {Tamfal}, {Mayer}, \& {Sesana}}]{2019MNRAS.487.4985B}
{Biava}, N., {Colpi}, M., {Capelo}, P.~R., {et~al.} 2019, \mnras, 487, 4985, \dodoi{10.1093/mnras/stz1614}

\bibitem[{{Bond} {et~al.}(1991){Bond}, {Cole}, {Efstathiou}, \& {Kaiser}}]{1991ApJ...379..440B}
{Bond}, J.~R., {Cole}, S., {Efstathiou}, G., \& {Kaiser}, N. 1991, \apj, 379, 440, \dodoi{10.1086/170520}

\bibitem[{{Bonetti} {et~al.}(2018){Bonetti}, {Sesana}, {Barausse}, \& {Haardt}}]{2018MNRAS.477.2599B}
{Bonetti}, M., {Sesana}, A., {Barausse}, E., \& {Haardt}, F. 2018, \mnras, 477, 2599, \dodoi{10.1093/mnras/sty874}

\bibitem[{{Campeti} {et~al.}(2021){Campeti}, {Komatsu}, {Poletti}, \& {Baccigalupi}}]{2021JCAP...01..012C}
{Campeti}, P., {Komatsu}, E., {Poletti}, D., \& {Baccigalupi}, C. 2021, \jcap, 2021, 012, \dodoi{10.1088/1475-7516/2021/01/012}

\bibitem[{{Chakraborty} {et~al.}(2023){Chakraborty}, {Gallerani}, {Zana}, {Sesana}, {Valentini}, {Izquierdo-Villalba}, {Di Mascia}, {Vito}, \& {Barai}}]{2023MNRAS.523..758C}
{Chakraborty}, S., {Gallerani}, S., {Zana}, T., {et~al.} 2023, \mnras, 523, 758, \dodoi{10.1093/mnras/stad1493}

\bibitem[{{Colpi}(2014)}]{2014SSRv..183..189C}
{Colpi}, M. 2014, \ssr, 183, 189, \dodoi{10.1007/s11214-014-0067-1}

\bibitem[{{Colpi} {et~al.}(1999){Colpi}, {Mayer}, \& {Governato}}]{1999ApJ...525..720C}
{Colpi}, M., {Mayer}, L., \& {Governato}, F. 1999, \apj, 525, 720, \dodoi{10.1086/307952}

\bibitem[{{Cury{\l}o} \& {Bulik}(2022)}]{2022A&A...660A..68C}
{Cury{\l}o}, M., \& {Bulik}, T. 2022, \aap, 660, A68, \dodoi{10.1051/0004-6361/202141987}

\bibitem[{{Dijkstra} {et~al.}(2008){Dijkstra}, {Haiman}, {Mesinger}, \& {Wyithe}}]{2008MNRAS.391.1961D}
{Dijkstra}, M., {Haiman}, Z., {Mesinger}, A., \& {Wyithe}, J. S.~B. 2008, \mnras, 391, 1961, \dodoi{10.1111/j.1365-2966.2008.14031.x}

\bibitem[{{Enoki} {et~al.}(2004){Enoki}, {Inoue}, {Nagashima}, \& {Sugiyama}}]{2004ApJ...615...19E}
{Enoki}, M., {Inoue}, K.~T., {Nagashima}, M., \& {Sugiyama}, N. 2004, \apj, 615, 19, \dodoi{10.1086/424475}

\bibitem[{{Enoki} \& {Nagashima}(2007)}]{2007PThPh.117..241E}
{Enoki}, M., \& {Nagashima}, M. 2007, Progress of Theoretical Physics, 117, 241, \dodoi{10.1143/PTP.117.241}

\bibitem[{{Ferrarese}(2002)}]{2002ApJ...578...90F}
{Ferrarese}, L. 2002, \apj, 578, 90, \dodoi{10.1086/342308}

\bibitem[{{Greif} {et~al.}(2011){Greif}, {Springel}, {White}, {Glover}, {Clark}, {Smith}, {Klessen}, \& {Bromm}}]{2011ApJ...737...75G}
{Greif}, T.~H., {Springel}, V., {White}, S. D.~M., {et~al.} 2011, \apj, 737, 75, \dodoi{10.1088/0004-637X/737/2/75}

\bibitem[{{Habouzit} {et~al.}(2016){Habouzit}, {Volonteri}, {Latif}, {Dubois}, \& {Peirani}}]{2016MNRAS.463..529H}
{Habouzit}, M., {Volonteri}, M., {Latif}, M., {Dubois}, Y., \& {Peirani}, S. 2016, \mnras, 463, 529, \dodoi{10.1093/mnras/stw1924}

\bibitem[{{Hobbs} {et~al.}(2010){Hobbs}, {Archibald}, {Arzoumanian}, {Backer}, {Bailes}, {Bhat}, {Burgay}, {Burke-Spolaor}, {Champion}, {Cognard}, {Coles}, {Cordes}, {Demorest}, {Desvignes}, {Ferdman}, {Finn}, {Freire}, {Gonzalez}, {Hessels}, {Hotan}, {Janssen}, {Jenet}, {Jessner}, {Jordan}, {Kaspi}, {Kramer}, {Kondratiev}, {Lazio}, {Lazaridis}, {Lee}, {Levin}, {Lommen}, {Lorimer}, {Lynch}, {Lyne}, {Manchester}, {McLaughlin}, {Nice}, {Oslowski}, {Pilia}, {Possenti}, {Purver}, {Ransom}, {Reynolds}, {Sanidas}, {Sarkissian}, {Sesana}, {Shannon}, {Siemens}, {Stairs}, {Stappers}, {Stinebring}, {Theureau}, {van Haasteren}, {van Straten}, {Verbiest}, {Yardley}, \& {You}}]{2010CQGra..27h4013H}
{Hobbs}, G., {Archibald}, A., {Arzoumanian}, Z., {et~al.} 2010, Classical and Quantum Gravity, 27, 084013, \dodoi{10.1088/0264-9381/27/8/084013}

\bibitem[{{Izquierdo-Villalba} {et~al.}(2022){Izquierdo-Villalba}, {Sesana}, {Bonoli}, \& {Colpi}}]{2022MNRAS.509.3488I}
{Izquierdo-Villalba}, D., {Sesana}, A., {Bonoli}, S., \& {Colpi}, M. 2022, \mnras, 509, 3488, \dodoi{10.1093/mnras/stab3239}

\bibitem[{{Jaffe} \& {Backer}(2003)}]{2003ApJ...583..616J}
{Jaffe}, A.~H., \& {Backer}, D.~C. 2003, \apj, 583, 616, \dodoi{10.1086/345443}

\bibitem[{{Katz} {et~al.}(2020){Katz}, {Kelley}, {Dosopoulou}, {Berry}, {Blecha}, \& {Larson}}]{2020MNRAS.491.2301K}
{Katz}, M.~L., {Kelley}, L.~Z., {Dosopoulou}, F., {et~al.} 2020, \mnras, 491, 2301, \dodoi{10.1093/mnras/stz3102}

\bibitem[{{Kelley} {et~al.}(2017){Kelley}, {Blecha}, {Hernquist}, {Sesana}, \& {Taylor}}]{2017MNRAS.471.4508K}
{Kelley}, L.~Z., {Blecha}, L., {Hernquist}, L., {Sesana}, A., \& {Taylor}, S.~R. 2017, \mnras, 471, 4508, \dodoi{10.1093/mnras/stx1638}

\bibitem[{{Khan} {et~al.}(2016){Khan}, {Husa}, {Hannam}, {Ohme}, {P{\"u}rrer}, {Forteza}, \& {Boh{\'e}}}]{2016PhRvD..93d4007K}
{Khan}, S., {Husa}, S., {Hannam}, M., {et~al.} 2016, \prd, 93, 044007, \dodoi{10.1103/PhysRevD.93.044007}

\bibitem[{{Kramer} \& {Champion}(2013)}]{2013CQGra..30v4009K}
{Kramer}, M., \& {Champion}, D.~J. 2013, Classical and Quantum Gravity, 30, 224009, \dodoi{10.1088/0264-9381/30/22/224009}

\bibitem[{{Kroupa} {et~al.}(2020){Kroupa}, {Subr}, {Jerabkova}, \& {Wang}}]{2020MNRAS.498.5652K}
{Kroupa}, P., {Subr}, L., {Jerabkova}, T., \& {Wang}, L. 2020, \mnras, 498, 5652, \dodoi{10.1093/mnras/staa2276}

\bibitem[{{Lacey} \& {Cole}(1993)}]{1993MNRAS.262..627L}
{Lacey}, C., \& {Cole}, S. 1993, \mnras, 262, 627, \dodoi{10.1093/mnras/262.3.627}

\bibitem[{Maggiore(2018)}]{Maggiore:2018sht}
Maggiore, M. 2018, {Gravitational Waves. Vol. 2: Astrophysics and Cosmology} (Oxford University Press)

\bibitem[{{Manchester} {et~al.}(2013){Manchester}, {Hobbs}, {Bailes}, {Coles}, {van Straten}, {Keith}, {Shannon}, {Bhat}, {Brown}, {Burke-Spolaor}, {Champion}, {Chaudhary}, {Edwards}, {Hampson}, {Hotan}, {Jameson}, {Jenet}, {Kesteven}, {Khoo}, {Kocz}, {Maciesiak}, {Oslowski}, {Ravi}, {Reynolds}, {Sarkissian}, {Verbiest}, {Wen}, {Wilson}, {Yardley}, {Yan}, \& {You}}]{2013PASA...30...17M}
{Manchester}, R.~N., {Hobbs}, G., {Bailes}, M., {et~al.} 2013, \pasa, 30, e017, \dodoi{10.1017/pasa.2012.017}

\bibitem[{{Marasco} {et~al.}(2021){Marasco}, {Cresci}, {Posti}, {Fraternali}, {Mannucci}, {Marconi}, {Belfiore}, \& {Fall}}]{2021MNRAS.507.4274M}
{Marasco}, A., {Cresci}, G., {Posti}, L., {et~al.} 2021, \mnras, 507, 4274, \dodoi{10.1093/mnras/stab2317}

\bibitem[{{Mayer}(2017)}]{2017JPhCS.840a2025M}
{Mayer}, L. 2017, in Journal of Physics Conference Series, Vol. 840, Journal of Physics Conference Series, 012025, \dodoi{10.1088/1742-6596/840/1/012025}

\bibitem[{{McConnell} \& {Ma}(2013)}]{2013ApJ...764..184M}
{McConnell}, N.~J., \& {Ma}, C.-P. 2013, \apj, 764, 184, \dodoi{10.1088/0004-637X/764/2/184}

\bibitem[{{McLaughlin}(2013)}]{2013CQGra..30v4008M}
{McLaughlin}, M.~A. 2013, Classical and Quantum Gravity, 30, 224008, \dodoi{10.1088/0264-9381/30/22/224008}

\bibitem[{{Ravi} {et~al.}(2015){Ravi}, {Wyithe}, {Shannon}, \& {Hobbs}}]{2015MNRAS.447.2772R}
{Ravi}, V., {Wyithe}, J.~S.~B., {Shannon}, R.~M., \& {Hobbs}, G. 2015, \mnras, 447, 2772, \dodoi{10.1093/mnras/stu2659}

\bibitem[{{Ravi} {et~al.}(2014){Ravi}, {Wyithe}, {Shannon}, {Hobbs}, \& {Manchester}}]{2014MNRAS.442...56R}
{Ravi}, V., {Wyithe}, J.~S.~B., {Shannon}, R.~M., {Hobbs}, G., \& {Manchester}, R.~N. 2014, \mnras, 442, 56, \dodoi{10.1093/mnras/stu779}

\bibitem[{{Reardon} {et~al.}(2023){Reardon}, {Zic}, {Shannon}, {Hobbs}, {Bailes}, {Di Marco}, {Kapur}, {Rogers}, {Thrane}, {Askew}, {Bhat}, {Cameron}, {Cury{\l}o}, {Coles}, {Dai}, {Goncharov}, {Kerr}, {Kulkarni}, {Levin}, {Lower}, {Manchester}, {Mandow}, {Miles}, {Nathan}, {Os{\l}owski}, {Russell}, {Spiewak}, {Zhang}, \& {Zhu}}]{2023ApJ...951L...6R}
{Reardon}, D.~J., {Zic}, A., {Shannon}, R.~M., {et~al.} 2023, \apjl, 951, L6, \dodoi{10.3847/2041-8213/acdd02}

\bibitem[{{Salcido} {et~al.}(2016){Salcido}, {Bower}, {Theuns}, {McAlpine}, {Schaller}, {Crain}, {Schaye}, \& {Regan}}]{2016MNRAS.463..870S}
{Salcido}, J., {Bower}, R.~G., {Theuns}, T., {et~al.} 2016, \mnras, 463, 870, \dodoi{10.1093/mnras/stw2048}

\bibitem[{{Schneider}(2006)}]{2006NewAR..50...64S}
{Schneider}, R. 2006, \nar, 50, 64, \dodoi{10.1016/j.newar.2005.11.014}

\bibitem[{{Sesana}(2013)}]{2013MNRAS.433L...1S}
{Sesana}, A. 2013, \mnras, 433, L1, \dodoi{10.1093/mnrasl/slt034}

\bibitem[{{Sesana} {et~al.}(2004){Sesana}, {Haardt}, {Madau}, \& {Volonteri}}]{2004ApJ...611..623S}
{Sesana}, A., {Haardt}, F., {Madau}, P., \& {Volonteri}, M. 2004, \apj, 611, 623, \dodoi{10.1086/422185}

\bibitem[{{Sesana} {et~al.}(2008){Sesana}, {Vecchio}, \& {Colacino}}]{2008MNRAS.390..192S}
{Sesana}, A., {Vecchio}, A., \& {Colacino}, C.~N. 2008, \mnras, 390, 192, \dodoi{10.1111/j.1365-2966.2008.13682.x}

\bibitem[{{Sesana} {et~al.}(2009){Sesana}, {Vecchio}, \& {Volonteri}}]{2009MNRAS.394.2255S}
{Sesana}, A., {Vecchio}, A., \& {Volonteri}, M. 2009, \mnras, 394, 2255, \dodoi{10.1111/j.1365-2966.2009.14499.x}

\bibitem[{{Shannon} {et~al.}(2015){Shannon}, {Ravi}, {Lentati}, {Lasky}, {Hobbs}, {Kerr}, {Manchester}, {Coles}, {Levin}, {Bailes}, {Bhat}, {Burke-Spolaor}, {Dai}, {Keith}, {Os{\l}owski}, {Reardon}, {van Straten}, {Toomey}, {Wang}, {Wen}, {Wyithe}, \& {Zhu}}]{2015Sci...349.1522S}
{Shannon}, R.~M., {Ravi}, V., {Lentati}, L.~T., {et~al.} 2015, Science, 349, 1522, \dodoi{10.1126/science.aab1910}

\bibitem[{{Shimasaku} \& {Izumi}(2019)}]{2019ApJ...872L..29S}
{Shimasaku}, K., \& {Izumi}, T. 2019, \apjl, 872, L29, \dodoi{10.3847/2041-8213/ab053f}

\bibitem[{{Somerville} \& {Kolatt}(1999)}]{1999MNRAS.305....1S}
{Somerville}, R.~S., \& {Kolatt}, T.~S. 1999, \mnras, 305, 1, \dodoi{10.1046/j.1365-8711.1999.02154.x}

\bibitem[{{Spinoso} {et~al.}(2023){Spinoso}, {Bonoli}, {Valiante}, {Schneider}, \& {Izquierdo-Villalba}}]{2023MNRAS.518.4672S}
{Spinoso}, D., {Bonoli}, S., {Valiante}, R., {Schneider}, R., \& {Izquierdo-Villalba}, D. 2023, \mnras, 518, 4672, \dodoi{10.1093/mnras/stac3169}

\bibitem[{{Taffoni} {et~al.}(2003){Taffoni}, {Mayer}, {Colpi}, \& {Governato}}]{2003MNRAS.341..434T}
{Taffoni}, G., {Mayer}, L., {Colpi}, M., \& {Governato}, F. 2003, \mnras, 341, 434, \dodoi{10.1046/j.1365-8711.2003.06395.x}

\bibitem[{{Trinca} {et~al.}(2022){Trinca}, {Schneider}, {Valiante}, {Graziani}, {Zappacosta}, \& {Shankar}}]{2022MNRAS.511..616T}
{Trinca}, A., {Schneider}, R., {Valiante}, R., {et~al.} 2022, \mnras, 511, 616, \dodoi{10.1093/mnras/stac062}

\bibitem[{{Valiante} {et~al.}(2016){Valiante}, {Schneider}, {Volonteri}, \& {Omukai}}]{2016MNRAS.457.3356V}
{Valiante}, R., {Schneider}, R., {Volonteri}, M., \& {Omukai}, K. 2016, \mnras, 457, 3356, \dodoi{10.1093/mnras/stw225}

\bibitem[{{Volonteri} {et~al.}(2003){Volonteri}, {Haardt}, \& {Madau}}]{2003ApJ...582..559V}
{Volonteri}, M., {Haardt}, F., \& {Madau}, P. 2003, \apj, 582, 559, \dodoi{10.1086/344675}

\bibitem[{{Volonteri} {et~al.}(2020){Volonteri}, {Pfister}, {Beckmann}, {Dubois}, {Colpi}, {Conselice}, {Dotti}, {Martin}, {Jackson}, {Kraljic}, {Pichon}, {Trebitsch}, {Yi}, {Devriendt}, \& {Peirani}}]{2020MNRAS.498.2219V}
{Volonteri}, M., {Pfister}, H., {Beckmann}, R.~S., {et~al.} 2020, \mnras, 498, 2219, \dodoi{10.1093/mnras/staa2384}

\bibitem[{{Wyithe} \& {Loeb}(2003)}]{2003ApJ...590..691W}
{Wyithe}, J. S.~B., \& {Loeb}, A. 2003, \apj, 590, 691, \dodoi{10.1086/375187}

\bibitem[{{Xu} {et~al.}(2023){Xu}, {Chen}, {Guo}, {Jiang}, {Wang}, {Xu}, {Xue}, {Nicolas Caballero}, {Yuan}, {Xu}, {Wang}, {Hao}, {Luo}, {Lee}, {Han}, {Jiang}, {Shen}, {Wang}, {Wang}, {Xu}, {Wu}, {Manchester}, {Qian}, {Guan}, {Huang}, {Sun}, \& {Zhu}}]{2023RAA....23g5024X}
{Xu}, H., {Chen}, S., {Guo}, Y., {et~al.} 2023, Research in Astronomy and Astrophysics, 23, 075024, \dodoi{10.1088/1674-4527/acdfa5}

\bibitem[{{Yang} {et~al.}(2019){Yang}, {Hu}, \& {Li}}]{2019MNRAS.483..503Y}
{Yang}, Q., {Hu}, B., \& {Li}, X.-D. 2019, \mnras, 483, 503, \dodoi{10.1093/mnras/sty3126}

\bibitem[{{Zhang} {et~al.}(2008){Zhang}, {Fakhouri}, \& {Ma}}]{2008MNRAS.389.1521Z}
{Zhang}, J., {Fakhouri}, O., \& {Ma}, C.-P. 2008, \mnras, 389, 1521, \dodoi{10.1111/j.1365-2966.2008.13671.x}

\end{thebibliography}
\bibliographystyle{aasjournal}

\end{document}